\documentclass{aa}  

\usepackage{graphicx}
\usepackage{txfonts}

\usepackage{natbib}
\bibpunct{(}{)}{;}{a}{}{,} 

\begin{document}

   \title{Planetary nebulae seen with TESS: Discovery of new binary central star candidates from Cycle 1}
   \titlerunning{Southern Planetary nebulae seen with TESS}

   \author{A. Aller
          \inst{1,2},
          J. Lillo-Box\inst{1}, 
           D. Jones\inst{3,4},
           L. F. Miranda\inst{5},    
            \and
           S. Barcel\'o Forteza\inst{1}       
          }

  \institute{Departamento de Astrof\'{\i}sica, Centro de Astrobiolog\'{\i}a
                (INTA-CSIC), ESAC Campus, Camino Bajo del Castillo s/n, 28692, Villanueva de la Ca\~nada (Madrid), Spain \\
              \email{alba.aller@cab.inta-csic.es}
                \and 
         Spanish Virtual Observatory, Spain\
         \and 
         Instituto de Astrof\'isica de Canarias, 38205 La Laguna, Tenerife, Spain\
         \and 
         Departamento de Astrof\'isica, Universidad de La Laguna, 38206 La Laguna, Tenerife, Spain\
          \and   
          Instituto de Astrof\'{\i}sica de Andaluc\'{\i}a - CSIC, C/ Glorieta de la Astronom\'{i}a s/n, 
             E-18008 Granada, Spain\
             }

 
  \abstract
   {It has become clear in recent years that binarity plays a crucial role in many aspects of planetary nebulae (PNe), particularly with regard to the striking morphologies they exhibit. To date, there are nearly 60 known binary central stars of PNe (bCSPNe). However, both theory and observation indicate that this figure represents only the tip of the iceberg, with the Galactic PN population hosting orders of magnitude more stars.}
   {We are involved in a search for new bCSPNe with the aim of enhancing the statistical validation of the key role of binarity in the formation and shaping of PNe. New discoveries of bCSPNe and their characterization carry important implications not only for understanding PN evolution, but also for studying binary evolution and the common-envelope phase, which is still poorly understood. }
   {We used data from the TESS satellite to search for variability in the eight CSPNe that belong to the two-minute cadence of preselected targets in Cycle 1, with their available pipeline-extracted light curves. We identified strong periodicities and analysed them in the context of the binary scenario.}
   {All the CSPNe but one (Abell\,15) show clear signs of periodic variability in TESS. The cause of this variability can be attributed to different effects, some of them requiring the presence of a companion star. We find simple sinusoidal modulations in several of the systems, compatible with irradiation effects. In addition, two of the central stars (PG\,1034+001 and NGC\,5189) also show photometric variations due to ellipsoidal variations and other signs of variability that are probably caused by star spots or relativistic Doppler-beaming. The case of the well-studied Helix Nebula is of particular interest; here we constructed a series of binary models to explain the modulations we see in the light curve. We find that the variability constrains the possible companion to be very low-mass main-sequence star or sub-stellar object. 
 We also identify, in substantial detail, the individual pulsation frequencies of NGC\,246.} 
   {}
  \maketitle
%

\section{Introduction}

In  recent years, a broad consensus has been reached on the role of central star binarity in explaining the departure from sphericity in planetary nebulae (PNe). Single-star scenarios can no longer explain the wide variety of striking morphologies shown by these stellar end products. There are more than 3000 PN identified in the Milky Way \citep{Frew-Parker2010} and roughly 80\% of them have complex morphologies which depart from sphericity; this includes, for example, elliptical, bipolar, and multipolar PNe, as well as other structures like jets, disks, and knots present in many of the PNe \citep[see e.g. ][]{Manchado1996a, Manchado1996b, Sahai-Trauger1998, Parker2006}. Since the discovery of the first binary central star of a planetary nebula (bCSPN) by \cite{Bond1976}, the total sample has been steadily growing. To date, there are fewer than 60 binary central stars known \citep[see ][]{Boffin-Jones2019}. However, the amazing variety of morphologies observed makes it highly likely that binaries are mainly responsible for the formation of a significant fraction (if not the majority) of non-spherical PNe. This means that the true population of bCSPNe should be far larger than what is currently known.

The orbital periods of the known sample of bCSPNe cluster between two hours and four days, approximately \citep{Boffin-Jones2019}. Just a few systems populate the longer period zone, although this is probably an observational bias since they are really difficult to detect through photometric variability. This is why most of the long period binary central stars have been discovered through radial velocity observations \citep[see e.g.][]{VanWinckel2014}.

   \begin{table*}[t]
\caption{Log of the TESS observations for our CSPNe sample. PN\,G designations, common names, TESS input catalogue (TIC), equatorial coordinates, TESS magnitude, and the sector and camera of the observation of each target are listed.}             
\label{table:1}      
\centering          
\begin{tabular}{l c cc  c c c c }     
\hline\hline       
PNG     &       Common name     &       TIC&            $\alpha$(2000.0) & $\delta$(2000.0) & TESS$_{mag}$ & Sector & Camera   \\ 
\hline      
  PN\,G\,118.8-74.7     & NGC\,246      &  3905338      &  00$^h$\,47$^m$\,03$\fs$3         & -11$^{\circ}$\,52$'$\,19$''$   &12.19& 3      & 1     \\
  PN\,G\,215.5-30.8     & Abell\,7              &  169305167    & 05$^h$\,03$^m$\,07$\fs$5    & -15$^{\circ}$\,36$'$\,23$''$    &15.52& 5  & 2     \\  
   PN\,G\,233.5-16.3    & Abell\,15     &  471013573    &  06$^h$\,27$^m$\,02$\fs$0         & -25$^{\circ}$\,22$'$\,50$''$   &15.90& 6      & 2     \\
   PN\,G\,219.2+07.5    &  RWT\,152     &  65145453     &  07$^h$\,29$^m$\,58$\fs$5         & -02$^{\circ}$\,06$'$\,38$''$   &13.27& 7      & 1     \\  
   PN\,G\,278.1-05.9    &NGC\,2867      &  387196603    &  09$^h$\,21$^m$\,25$\fs$4         & -58$^{\circ}$\,18$'$\,41$''$   & 12.3& 9, 10  & 3     1\\
                                &PG\,1034+001& 124598476        & 10$^h$\,37$^m$\,03$\fs$8    & -00$^{\circ}$\,08$'$\,19$''$   &13.62& 9   & 1      \\
   PN\,G\,307.2-03.4    &NGC\,5189      &  341689253    & 13$^h$\,33$^m$\,32$\fs$9    & -65$^{\circ}$\,58$'$\,27$''$    & 14.27& 11        & 2     \\
   PN\,G\,18.8-74.7     &NGC\,7293      &  69813909     & 22$^h$\,29$^m$\,38$\fs$5    & -20$^{\circ}$\,50$'$\,14$''$    &13.93& 2  & 1     \\
\hline                  
\end{tabular}
\end{table*}

With regard to the close binary fraction (i.e.\ the fraction of PNe hosting binaries with periods between a few hours and a few days), recent surveys have shown that they represent about 12-21\% \citep{miszalski2009a} of the CSPNe, although this number may be a lower limit, especially given the observational challenges involved in constraining this fraction. Most of the known bCSPNe have been detected through modulations in the light curve \citep[see e.g.][]{bond2000, miszalski2009a}, revealing the presence of close companions with very short orbital periods \citep{miszalski2009a,miszalski2011}. Apart from eclipsing binaries, typical photometric modulations include irradiation effects (i.e. the reflection of the light of the hot primary by the less massive companion due to the large temperature difference between them), ellipsoidal modulations (i.e. tidal distortions due to the proximity of both stars) and Doppler beaming effects (which are the photometric imprints of the Doppler effect). These short-period binary systems must have undergone a common-envelope (CE) phase \citep{iben-livio1993}, during which the primary fills its Roche lobe and transfers mass to its companion in an unstable way, forming a CE within which the two components spiral-in towards each other, resulting in a much closer binary system and the ejection of the CE. This process culminates in the formation of a (most likely aspherical) PN.

Space missions like NASA's Transiting Exoplanet Survey Satellite \citep[TESS;][]{Ricker2015} provide a unique opportunity to detect new binary central stars. TESS was launched on 18 April 2018 and its science operations were begun after a three-month commissioning period on 25 July 2018. It is a two-year, all-sky photometric survey charged with the main goal of obtaining high-precision photometry of more than 200,000 selected bright stars with a cadence of about two minutes, observing in 26 different sky sectors. With four cameras observing each TESS data sector, the resulting field-of-view is 24x96 degrees, which allows it to obtain photometry of any target within this wide field. These full frame images (FFIs) are read out approximately every 30 minutes, establishing the cadence of the photometry for the targets in the FFIs. As with Kepler, the high-precision, high-cadence obeservations and the continuous photometric data that TESS provides in a long time span (in comparison with the limitations that we have from the ground) make TESS an ideal machine for searching for bCSPNe.

In this paper, we  use TESS observations to find new candidate short-period binaries in the nuclei of PNe. Even though the initial sample is not very large (only eight central stars were observed in the short-cadence mode in Cycle 1), we detected significant photometric variability for seven targets, that is, more than 80\% of the sample. In Section 2, we describe the PNe observed by TESS, the TESS observations, and the analysis of the data. In Section 3, we discuss the variability and periodicities of each central star and we conclude in Section 4 with some final remarks.
   

\section{TESS photometry}

\subsection{The PN sample}

 In its first year of observations, TESS observed the southern ecliptic hemisphere in a total of 13 sectors. This includes roughly half of the CSPNe in our Galaxy. However, only eight of them are among the targets observed at the two-minute cadence and, therefore, their light curves have already been published in the archive. These eight PNe are summarised in Table~1. The rest of the southern PNe observed by TESS are available in the FFIs, with a longer cadence of 30 minutes. They do not have light curves published in the archive, but their photometry can be obtained directly from the images. We are currently in the process of extracting the light curves from the FFIs in order to obtain a complete dataset for all the CSPNe, the results of which will be presented in a future publication.

\subsection{Description of the TESS data}
 \label{Sect:description_TESS_data}

As mentioned above, TESS is equipped with four identical cameras covering each sector. Each of these four cameras has four 2k$\times$2k CCDs with a pixel scale of 21 arcsec\,pixel$^{-1}$. The detector bandpass covers the spectrum from 600 to 1000 nm and is centred on 786.5 nm, just as for the traditional Cousins I-band. 
We used the light curves processed by the Science Processing Operations Center (SPOC) pipeline \citep{Jenkins2016} and retrieved them from the Mikulski Archive for Space Telescopes (MAST\footnote{\url{https://mast.stsci.edu/portal/Mashup/Clients/Mast/Portal.html}}). SPOC light curves contain both the SAP (Simple Aperture Photometry) flux and the Pre-search Data Conditioning SAP (PDCSAP) flux. In the PDCSAP, long-term trends have been removed from the fluxes using the co-trending basis vectors. For the purpose of this work, we used the PDC light curves without any additional treatment of the data, only removing those data points with a non-zero 'quality' flag since they could be affected by some anomalies, such as cosmic ray events or instrumental issues (see Sect. 9 in the TESS Science Data Products Description Document\footnote{\url{https://archive.stsci.edu/missions/tess/doc/EXP-TESS-ARC-ICD-TM-0014.pdf}}). 
 
\begin{figure*}
     \includegraphics[width=0.335\textwidth]{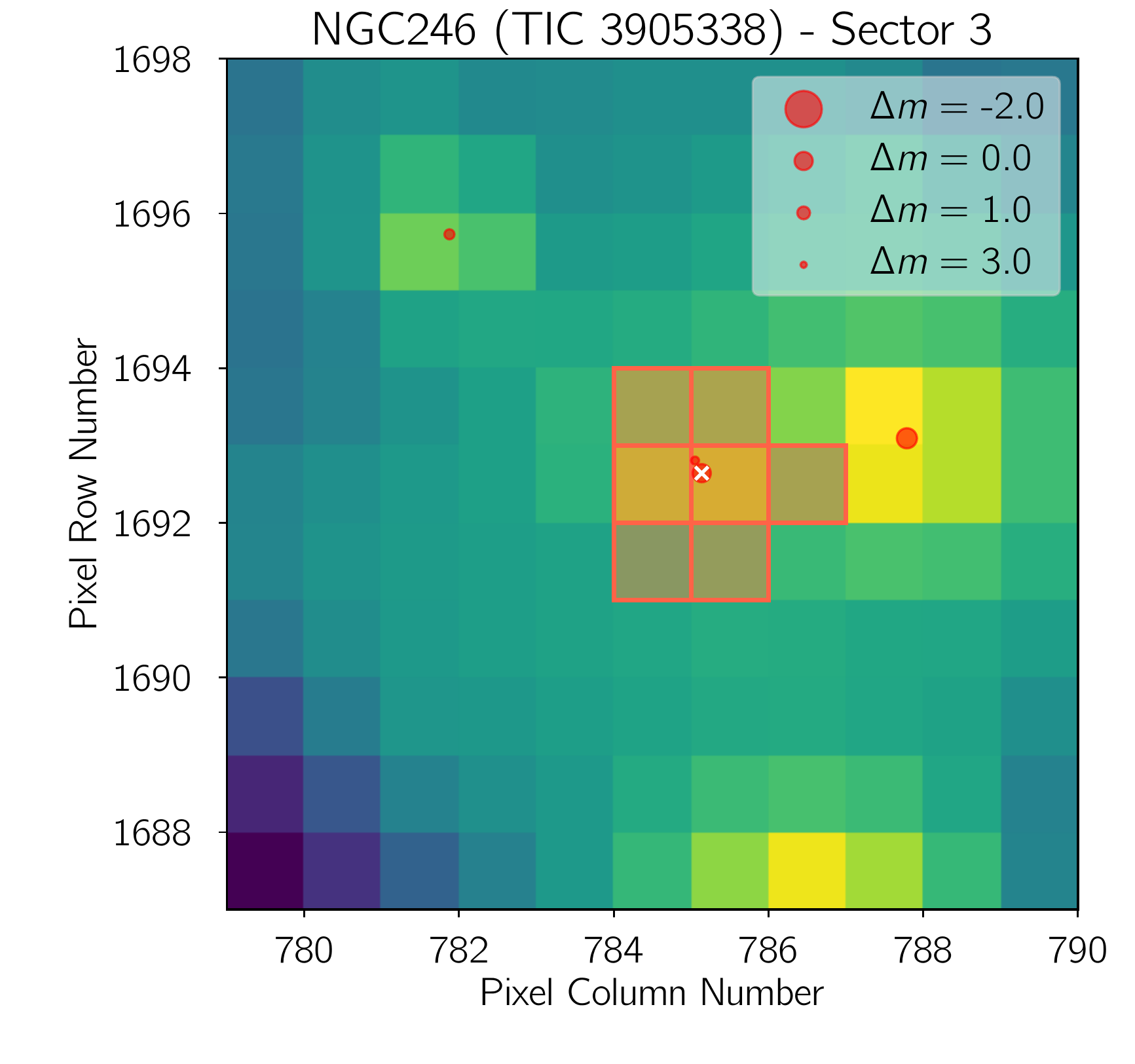}
     \includegraphics[width=0.335\textwidth]{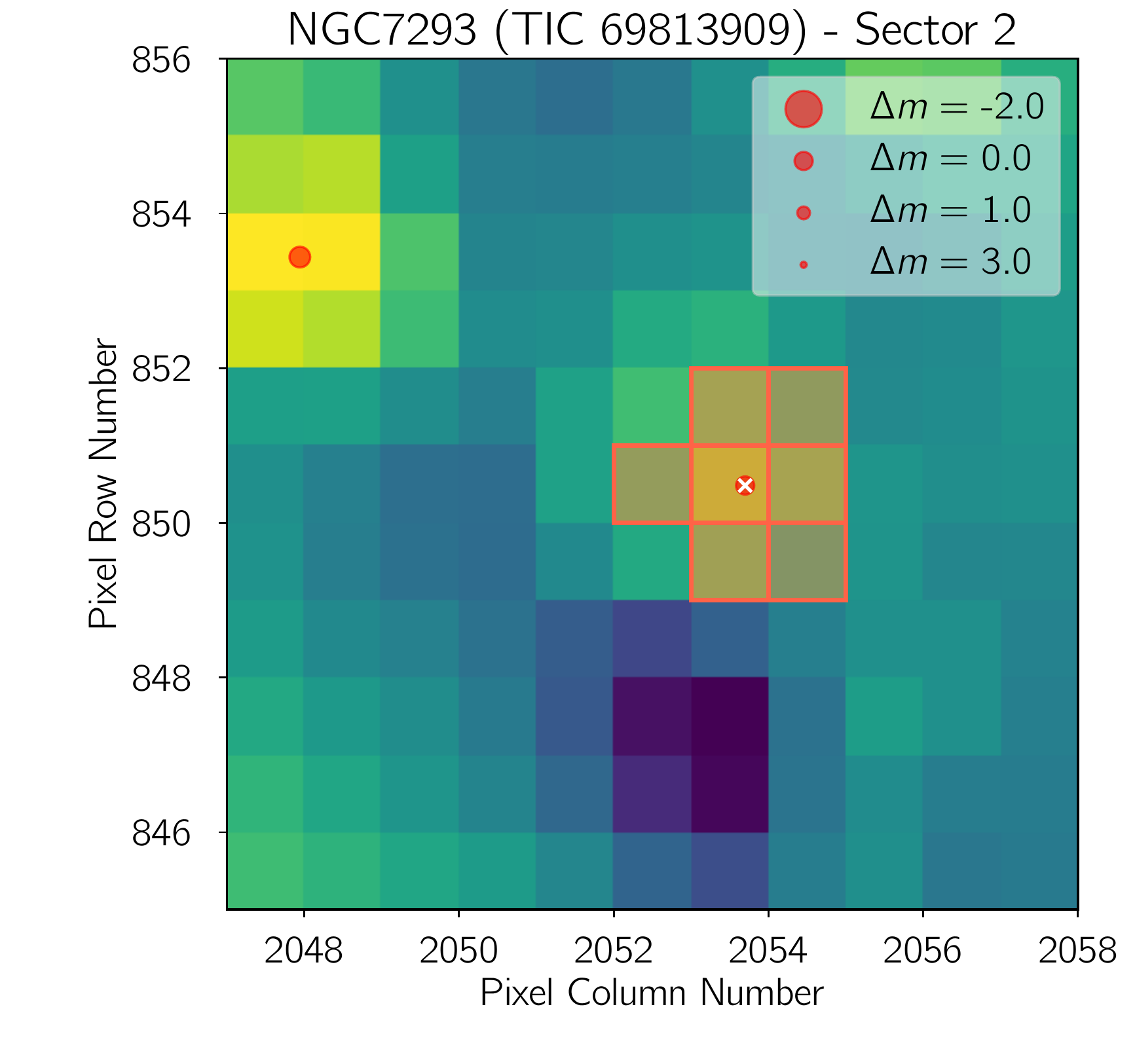}     
      \includegraphics[width=0.335\textwidth]{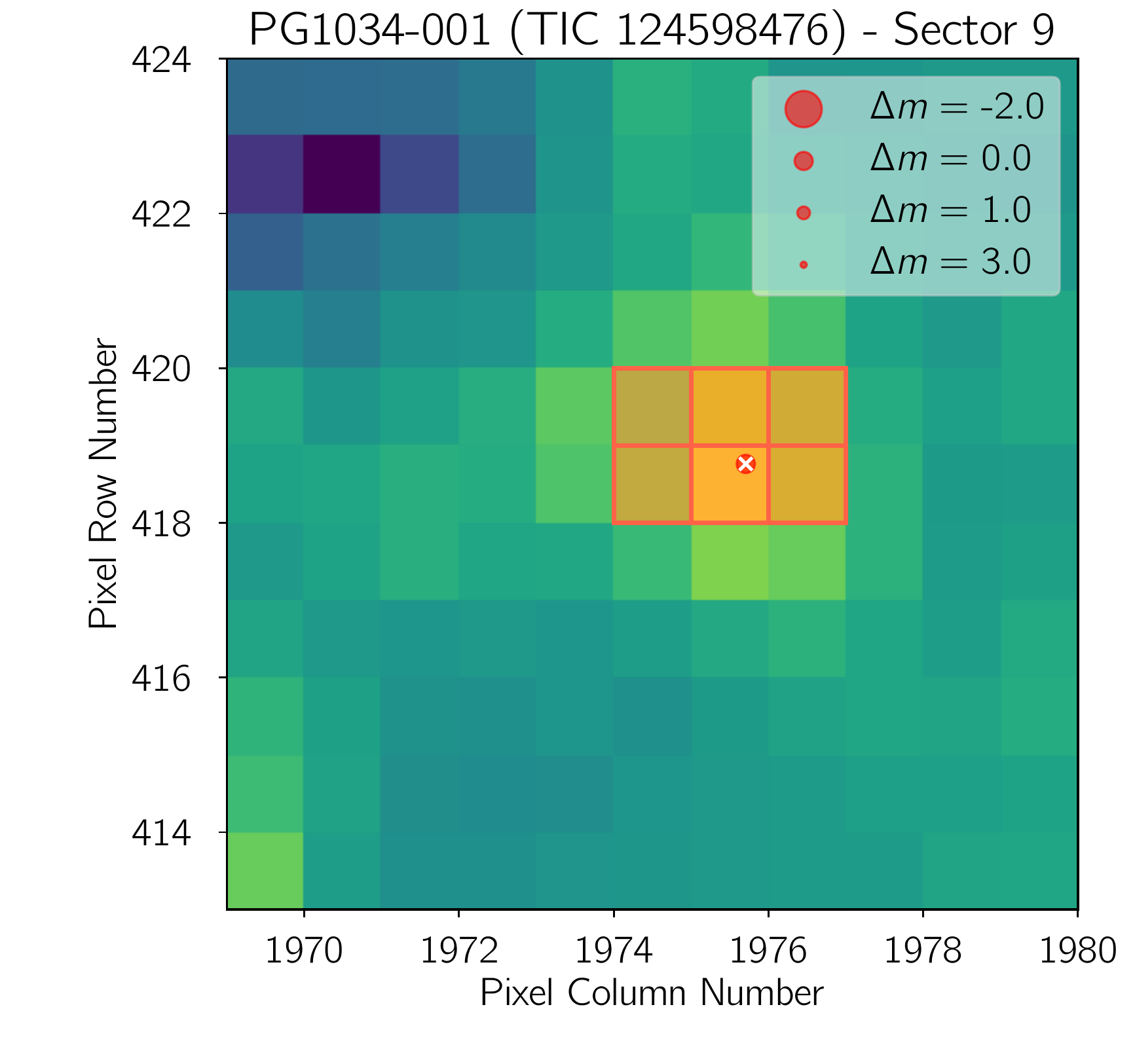}
       \includegraphics[width=0.335\textwidth]{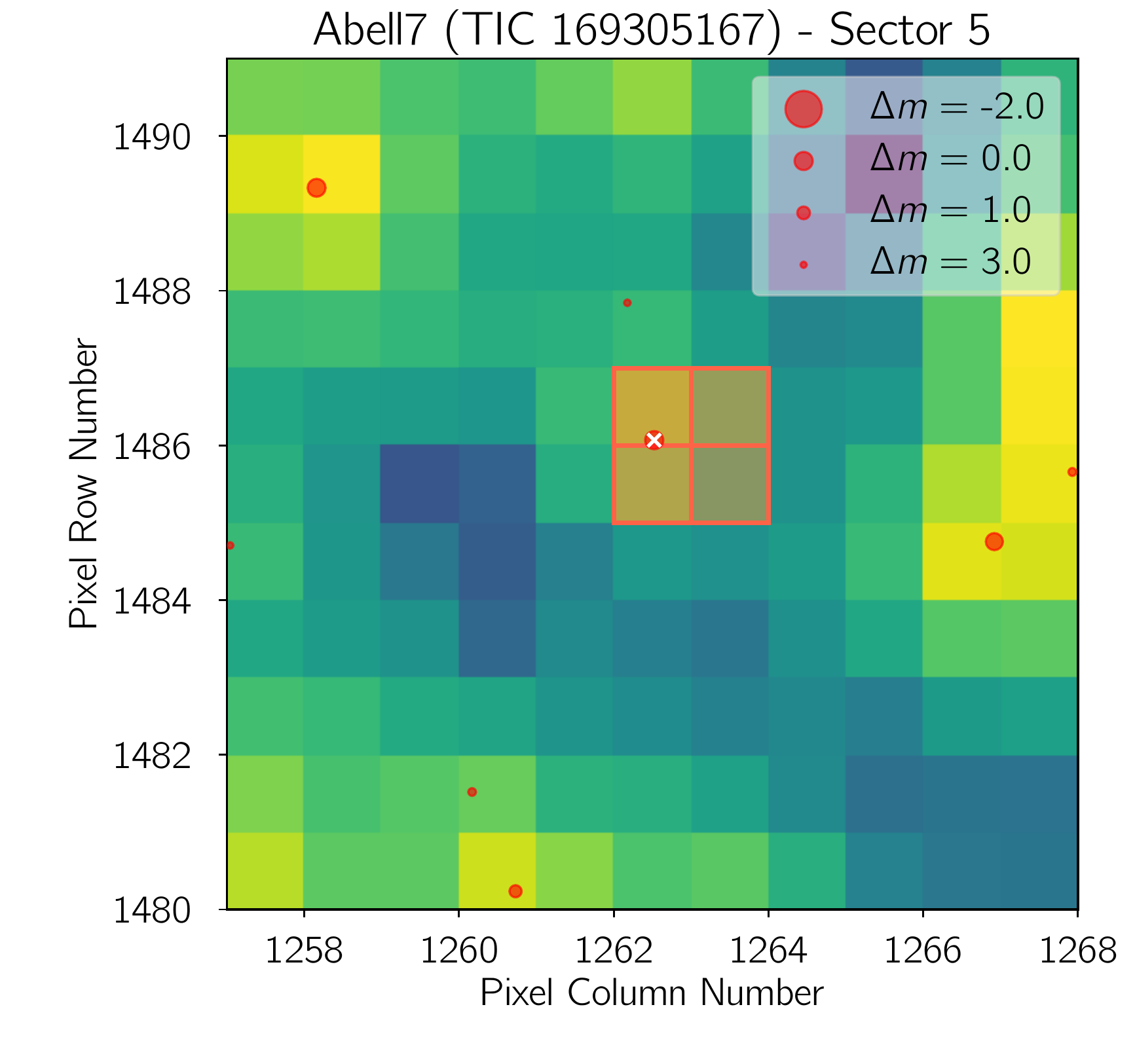}
       \includegraphics[width=0.335\textwidth]{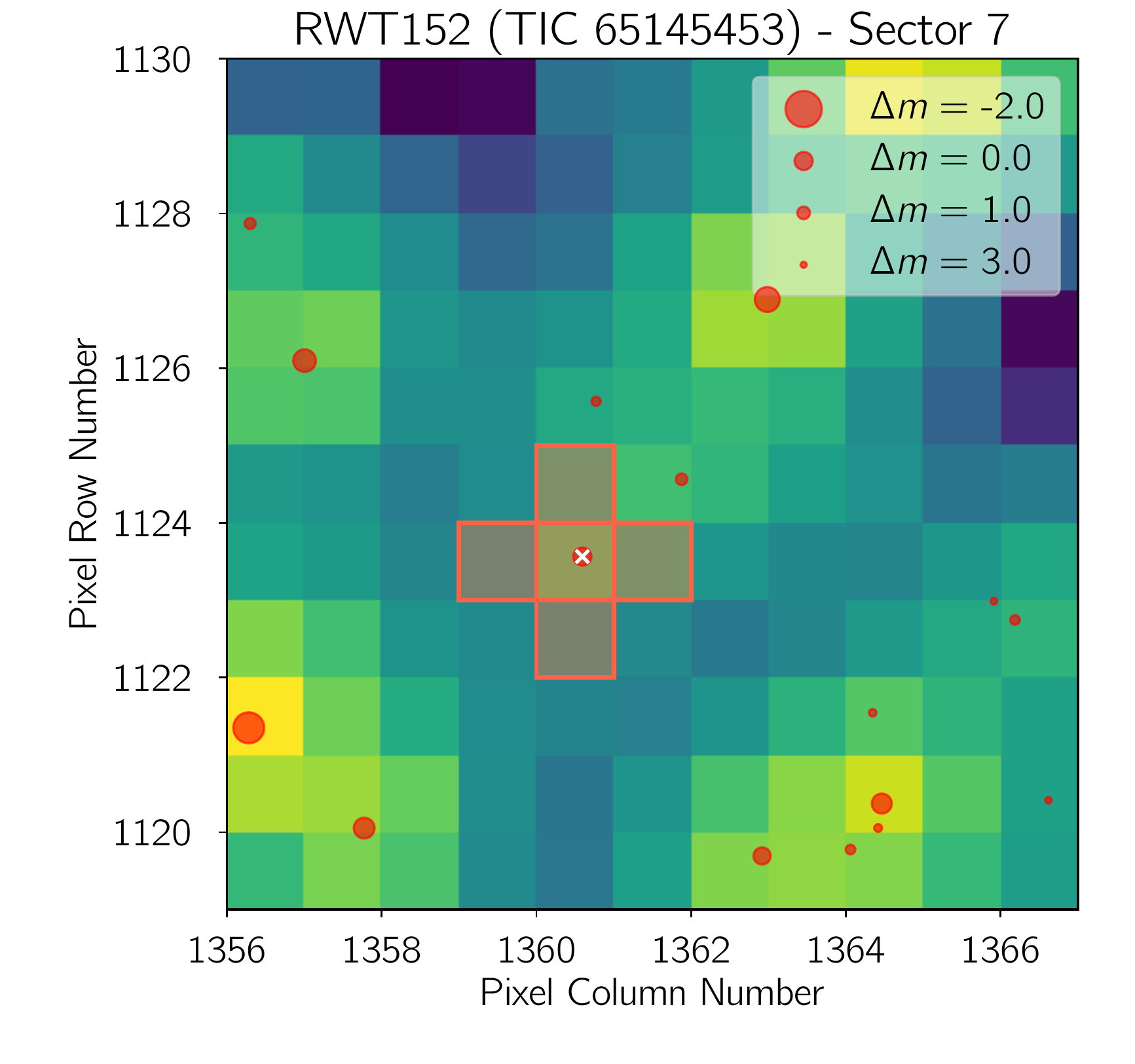}
        \includegraphics[width=0.335\textwidth]{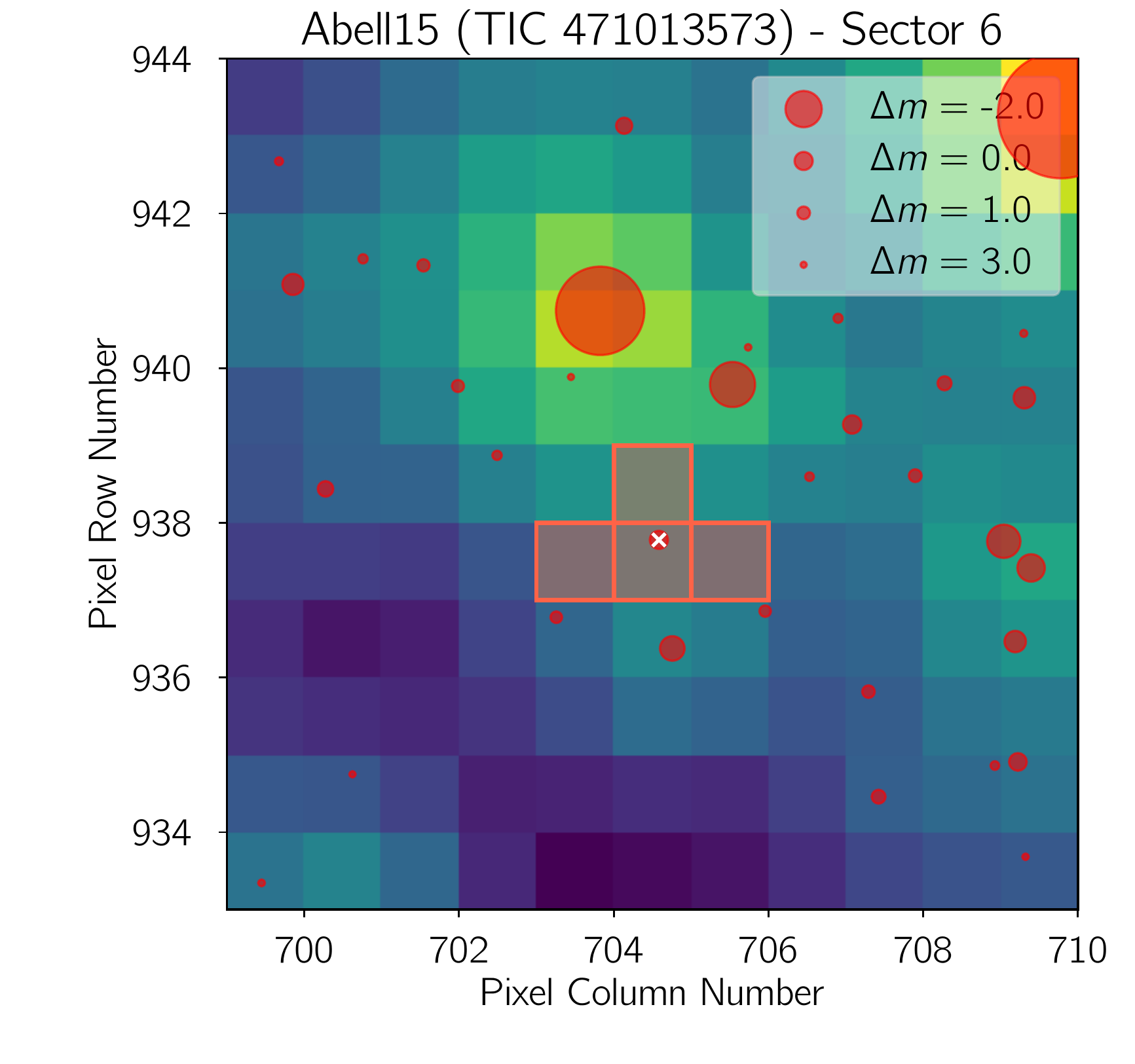}     
           \includegraphics[width=0.335\textwidth]{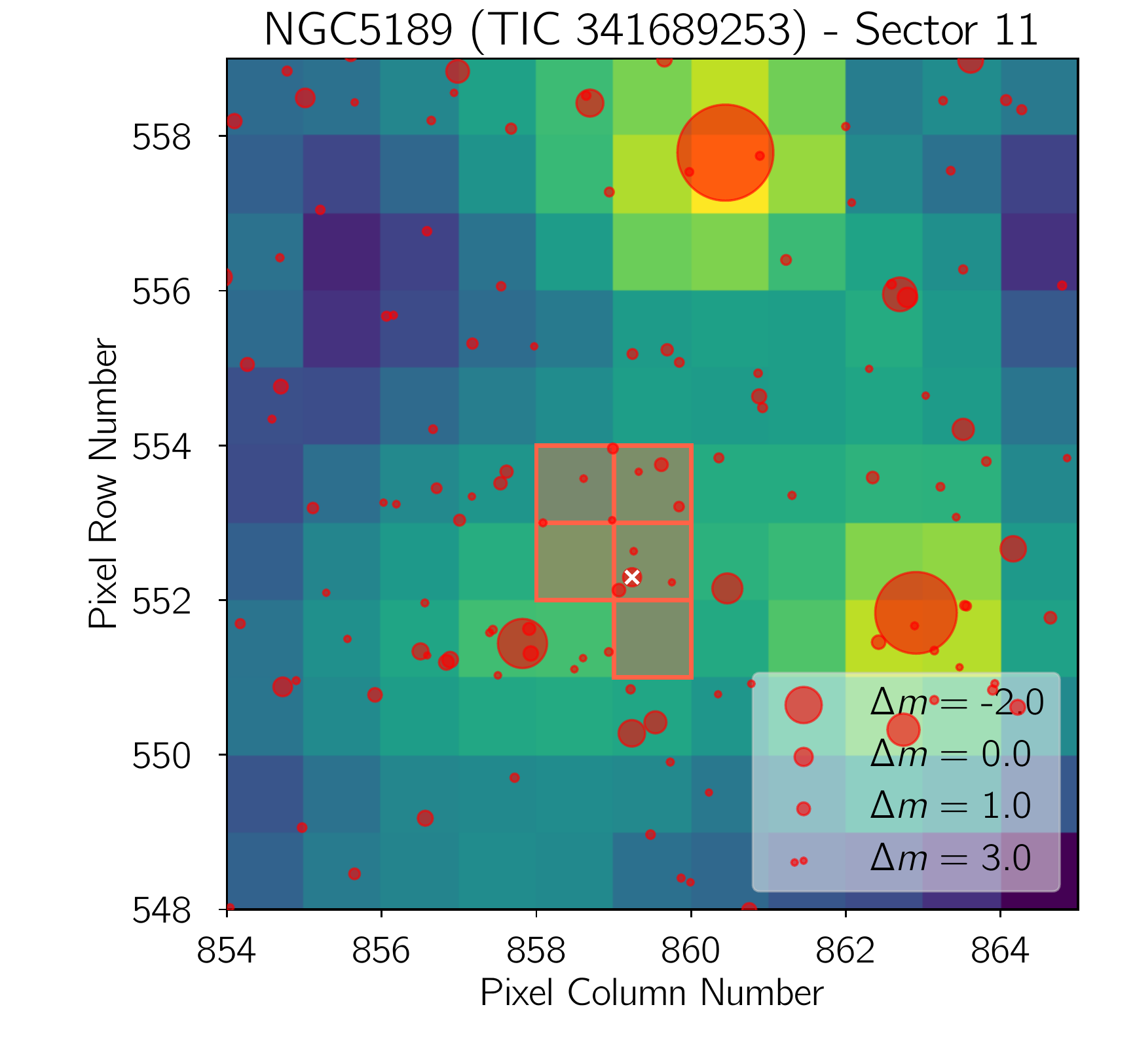}
          \includegraphics[width=0.335\textwidth]{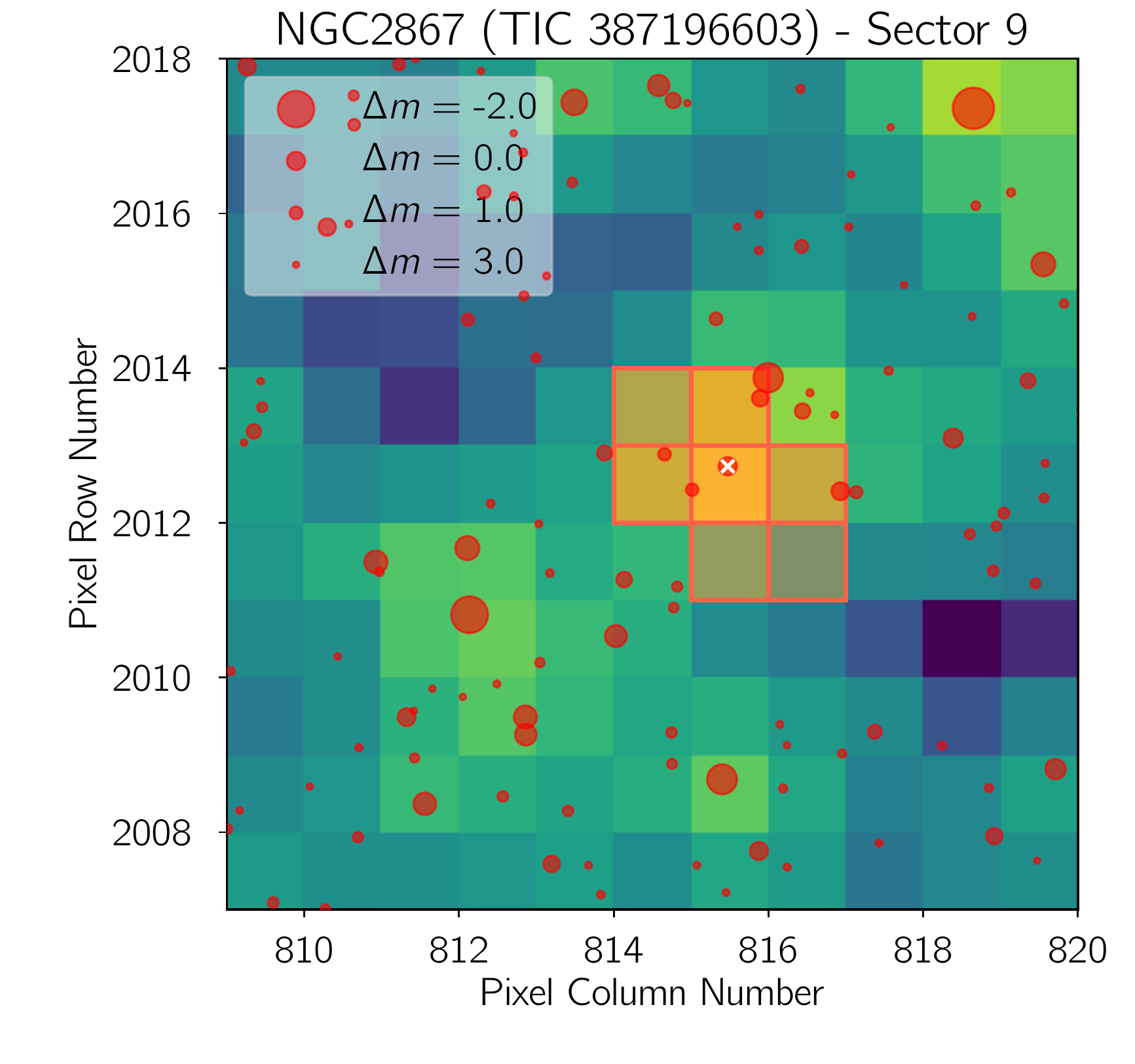}
          \includegraphics[width=0.335\textwidth]{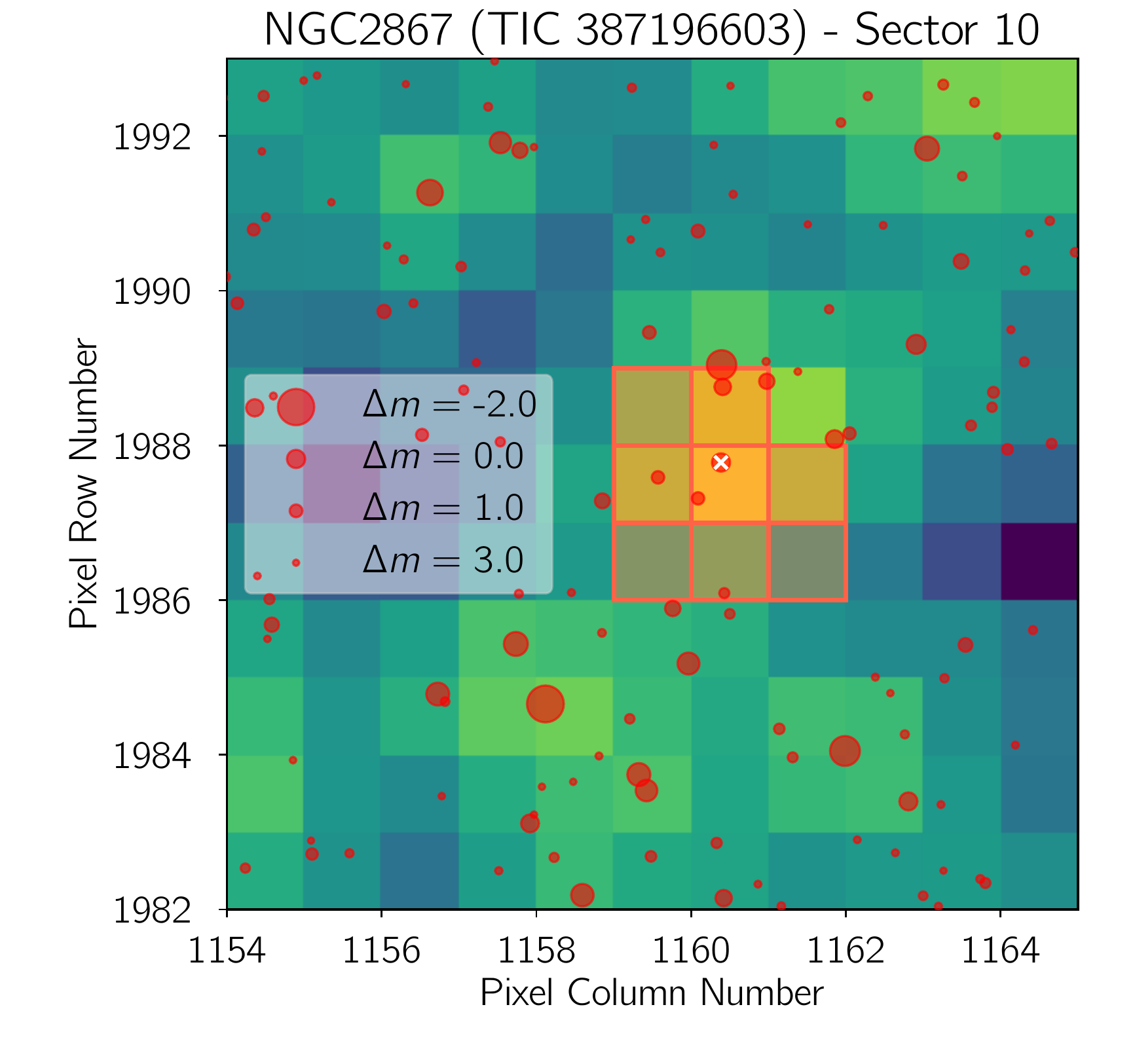}
\caption{Target pixel files (TPFs) of each central star in the sample. The red circles are the sources of the Gaia DR2 catalogue in the field with scaled magnitudes (see legend). White crosses indicate the location of the targets. The aperture mask used by the pipeline to extract the photometry is also marked. Pixel scale is 21 arcsec\,pixel$^{-1}$.}
\label{fig:TPFs}
\end{figure*}

We downloaded the target pixel file (TPF) of each object to inspect the aperture and possible contamination sources. This check of the TPFs is especially important because of the faintness of our sample in comparison with most of the two-minute cadences targets observed by TESS. Fig.~\ref{fig:TPFs} shows the TPF of each of the eight central stars in the sample, marked with a white cross. The red circles represent the Gaia sources from the DR2 catalogue \citep{Gaia2018}, scaled by magnitude contrast against the target source. We over-plotted all the sources with a magnitude contrast up to $\Delta$m = 3 (i.e. three magnitudes fainter than our target, which corresponds to $\sim$ 6\% of contamination if entirely inside the aperture). The aperture mask used by the pipeline to extract the photometry was also plotted over the TPF. In five cases (NGC\,7293, PG\,1034+001, Abell\,7, Abell\,15, and RWT\,152),  there is no contamination from other sources in the aperture mask. In the case of NGC\,246 there is minimal contamination by an object of Gaia magnitude 14.2 (2.4 magnitudes fainter than our target). Finally, NGC\,5189 and NGC\,2867 have severe contamination by several sources of similar magnitudes to the central star, which makes the photometry difficult to interpret. We will discuss these cases in detail in Section~\ref{Sect:analysis}.

\subsection{Data analysis and modelling}   
  
A Lomb-Scargle analysis was performed for each object in order to identify possible periodic variability in the data. This first inspection resulted in the detection of strong periodicities in all but one of the targets (Abell 15). We analysed the data in the time-flux domain with a simple sinusoidal model that includes the amplitude (A), the zero level ($Z_0$), period (P), and the time of conjunction ($T_0$) as parameters. In this first analysis, we assumed uniform priors on all of the parameters, namely $P\in\mathcal{U}(0.01,10)$ days, $A\in\mathcal{U}(0,40)$ parts per thousand (ppt), $T_0\in\mathcal{U}(t_{\rm min},t_{\rm max})$ days, and $Z_0\in\mathcal{U}(0.9,1.0)$. 

In order to correctly sample the posterior probability distribution of each of these parameters, we used the implementation of Goodman \& Weare's affine invariant Markov chain Monte Carlo (MCMC) ensemble sampler \textit{emcee}\footnote{See \url{http://dan.iel.fm/emcee/current/} for further documentation.}, developed by \cite{Foreman-mackey2013}. We used a first phase with 30 walkers and 5\,000 steps to find a first guess for the solution. Then we re-sampled the position of the walkers in a small N-dimensional ball around the best walker and ran a second phase with the same number of walkers and 2500 steps. All steps from the latter phase were used to compute the marginalised posterior distributions of each parameter  (no thinning was applied). The final chains are thus composed of 75\,000 steps that were used to compute the marginalised posterior probabilities for each parameter.

In the case of Abell\,7, two significant periodicities appear in the periodogram. The MCMC is not able to distinguish between these two similar periods. So we ran two separate MCMCs with Gaussian priors centered on each of the two periodicities ($P_1=$ 3.0 d, and $P_2=$2.6 d, see Sect.\ref{Sect: Abell7}). The periodogram of the residuals indicate that after removing one of the sinusoidal signals, the other periodicity still remains above the 0.1\% FAP (False Alarm Probability). Hence, we conclude that both frequencies are independent and probably come from different sources.

In two cases (PG\,1034+001 and NGC\,5189), the simple sinusoidal model did not produce a successful fit because of the presence of a two uneven maxima in the phase-folded light curves. We interpret this as a combination of ellipsoidal variations, Doppler beaming, and reflection. We then proceeded with a new model including all these three effects with the shape:

\begin{equation}
\frac{\Delta F}{F} = Z_{0} - A_{\rm ell} \cos{2\theta} - A_{\rm ref} \cos{\theta} + A_{\rm beam}\sin{\theta},
\end{equation}

\noindent where $\theta = 2\pi(t-T_0)/P)$. The amplitudes of the ellipsoidal, Doppler beaming, and reflection effects are represented by the parameters $A_{\rm ell}$, $A_{\rm bea}$, and $A_{\rm ref}$, respectively. It is important to note, however, that the model aims at representing the phase dependency for the three effects (ellipsoidal beaming and reflection) but does not put physical constraints on the physical parameters. Hence, it must be taken as the combination of sinusoidal functions with maxima at phases 0.25, 0.5, and 0.75. We proceeded in the same manner as explained above to obtain the posterior distributions of the parameters involved in this new model and obtained successful evidence against the simple sinusoidal model.

After applying the appropriate model to the each system, all the chains converged to the parameters shown in Table~\ref{Table:results_LCs} (for the objects showing only a simple sinusoidal) and Table~\ref{Table:results_LCs2} (for the cases of PG\,1034+001 and NGC\,5189, for which the three different effects have been modelled). These results are discussed in detail in Sect.~\ref{Sect:analysis}.

It is important to note that applying a purely sinusoidal model to the effects described can only be considered a rough estimation as the true light curve morphologies are a complex function of the stellar and systemic parameters (gravity and limb darkening, orbital inclination, etc.).

The precision of the TESS data also allows us to estimate the radial velocity amplitude of the stars in which a Doppler beaming factor is (a priori) present \citep[as explained by][for the case of KOI-74]{vanKerkwijk2010}. Thus, following the equations (2) and (3) from \cite{vanKerkwijk2010}, we can calculate the variation of the photon rate $n{_\gamma}$ observed by a telescope as follows:

  \begin{eqnarray}
           \frac{\Delta n{_\gamma}}{n{_\gamma}} = f_{\rm DB}  \frac{\varv{_r}}{c}      & = & - \frac{(hc/\lambda k T)e^{hc/\lambda k T}}{e^{hc/\lambda k T}-1} \frac{\varv{_r}}{c}\,
   ,\end{eqnarray}
   where $f_{\rm DB}$ is the Doppler boost pre-factor, $\lambda$ = 7865\,$\AA$ from TESS bandpass and $T$ is the temperature of the primary. As such, we can obtain an estimation for the binary mass function. We note that the functional form of the equations of \cite{vanKerkwijk2010} assumes that one component of the binary completely dominates the observed flux from the system (as is the case for KOI-74).  In systems where the stars have similar luminosities, the Doppler beaming effect of both stars leads to a sum that is destructive.

Figs.~\ref{fig:lightcurves_and_periodogram1},~\ref{fig:lightcurves_and_periodogram2},~\ref{fig:lightcurves_and_periodogram3},~\ref{fig:lightcurves_and_periodogram4}, and ~\ref{fig:lightcurves_and_periodogram5} show the time-flux light curve (top panels), the Lomb-Scargle periodograms (right panels) as well as the phase-folded light curves of each central star with the period derived in the fitting (left panels). The FAPs at 10\%, 1\%, and 0.1\% are indicated in each periodogram with grey horizontal dotted lines. Also, we have over-plotted in blue the periodogram of the residuals. For the purpose of visualisation, weighted means are employed to obtain the binned fluxes presented in the phase folded light curves.  Bin sizes were chosen to ensure 20 and 100 datapoints per orbit for the red and grey datapoints shown, respectively.

\section{Variability analysis and discussion}
\label{Sect:analysis}
In the following Section we describe the analysis of each central star individually and discuss the phase variations.

  \subsection{NGC\,7293}

Widely known as Helix nebula, NGC\,7293 is the nearest PN to us and possibly one of the most well-studied in the literature, with more than 800 references (most of these related to its amazing and complex morphology). It has been well-characterised thanks to numerous observations from a wide range of wavelengths and telescopes. It is a double ring PN, with an inner thick disk and multitude of cometary knots present in the nebular centre \citep[see e.g.][and references therein]{Meaburn1998, Henry1999, Odell2005, VandeSteene2015}. In addition, the whole nebula is surrounded by a massive molecular envelope \citep{Young1999}. The central nucleus is a hot \citep[T$_{\rm eff}$ = 100000 K;][]{Napiwotzki1999} DAO white dwarf with unexpected hard X-ray emission \citep[see e.g.][]{Leahy1994, Guerrero2001}. The origin of this emission is suggested to be an unseen dMe (dwarf M star with emission lines) companion, which would be also consistent with the variable H$\alpha$ emission line profile found by \cite{Gruendl2001}. However, there has beeen no evidence (until now) for such a companion in spectroscopic or photometric observations, nor in the HST imaging. Also, \citealt{Su2007} detected a dust disk around the central star by means of Spitzer observations. This disk was interpreted as a debris disk, although it differs substantially from the typical dust disc found around other white dwarfs.

The TESS light curve (see Fig.~\ref{fig:lightcurves_and_periodogram1}, first panel) shows for the first time a clear photometric variability with 2.77\,d period. The signal in the periodogram is very strong and above the 0.1\% FAP. As mentioned in Sect.~\ref{Sect:description_TESS_data}, there is no contamination in the aperture photometry by other sources in the field (within a magnitude limit of $\Delta$m = 3) which means that the variability is real and definitively comes from NGC\,7293. The light curve phase-folded with the 2.77\,d period (Fig.~\ref{fig:lightcurves_and_periodogram1}, first left panel) shows an amplitude of 1.7 ppt (see also Table~\ref{Table:results_LCs}; 1 ppt equals to 1.086 mmag). If the variability were attributable to ellipsoidal modulations, then the true period would be double and the light curve would display two minima per cycle (at each opposition).  However, this interpretation is unlikely given that the period would imply that the size of the tidally distorted star would be rather large, inconsistent with the measured surface gravity of the hot central star \citep{Napiwotzki1999} or a secondary with low enough luminosity to have escaped detection in previous studies.  Perhaps a more reasonable interpretation of the variability is that it originates from the irradiation of a cool companion \citep[e.g.][]{horvat19}.  In order to explore this hypothesis, we constructed a series of models using the \textsc{phoebe 2} code \citep{prsa16,jones2019} assuming that the mass, temperature, and radius of the primary corresponds to what is determined in \citet{Napiwotzki1999} and that the secondary is a main sequence star with mass, temperature, and radius relying on the empirical relations of \citet{Eker2018}. The primary was assumed to be a blackbody with limb-darkening coefficients extrapolated from \citet{gianninas13}, while the main sequence secondaries were modelled using \textsc{phoenix} atmospheres \citep{husser13} and interpolated limb-darkening as described in \citet{prsa16}.  The inclination of the binary was fixed based on the inclination of the nebula $i=28^\circ$ \citep{Henry1999} as is expected and observed in all systems for which both inclinations are known \citep{hillwig16}.  

We did not encounter any satisfactory fits for secondaries with masses in the range of $0.16 M_\odot \leq M_2 \leq 2.5 M_\odot$ (the range covered by the model atmospheres and empirical mass-radius-temperature relations used, which do not imply a secondary so bright that it would likely have been discovered by previous studies), with all models producing significantly greater variability amplitudes than those observed.  Even adjusting the parameters of the system, for example using the more modern evolutionary tracks of \citet{millerbertolami16}, which imply a lower mass and smaller radius than those used by \citet{Napiwotzki1999} or the lower inclination of \citet[][$i\sim21^\circ$]{odell98}, a satisfactory fit could not be found. This leaves a number of possible solutions: 1) The variability may be due to irradiation of a smaller body (i.e. a sub-stellar or planetary mass object) or 2) the variability may not be due to irradiation. In this case, the variability could be due to Doppler beaming \citep[][see Sect.~\ref{pg1034+001} for more details]{vanKerkwijk2010}, or perhaps due to spots on a rotating star. However, the scenario in which the Doppler beaming is the cause of the variability could be practically discarded, since the radial velocity amplitude derived following eq. (2) ($\varv{_r}$ $\approx$ 460 km s$^{-1}$) would involve a mass function of $f$(m) = ($P_{\rm orb}$/2$\pi$G) $\varv{_r}$ =  (m$_{2}$sin $i$)$^3$/(m$_{1}$+m$_{2}$)$^2$ = 28 M$_{\sun}$, which is highly unlikely. The clearest test of these possibilities will be high-resolution time-series spectroscopy, capable of differentiating the radial velocity signals that would be associated with these hypotheses.

\subsection{RWT\,152}

RWT\,152 is a very faint, bipolar planetary nebula with an sdO in its nucleus \citep{Aller2016}. The morphology and the displacement of the
central star with respect to the nebula's centre may indicate the presence of a possible binary system. From the high-resolution, long-slit spectra, \cite{Aller2015a} derived an inclination of 84 degrees for the equatorial plane of the nebula, which implies that the possible bCSPN (if any) would be almost edge-on. 

\begin{figure*}
       \includegraphics[width=0.99\textwidth]{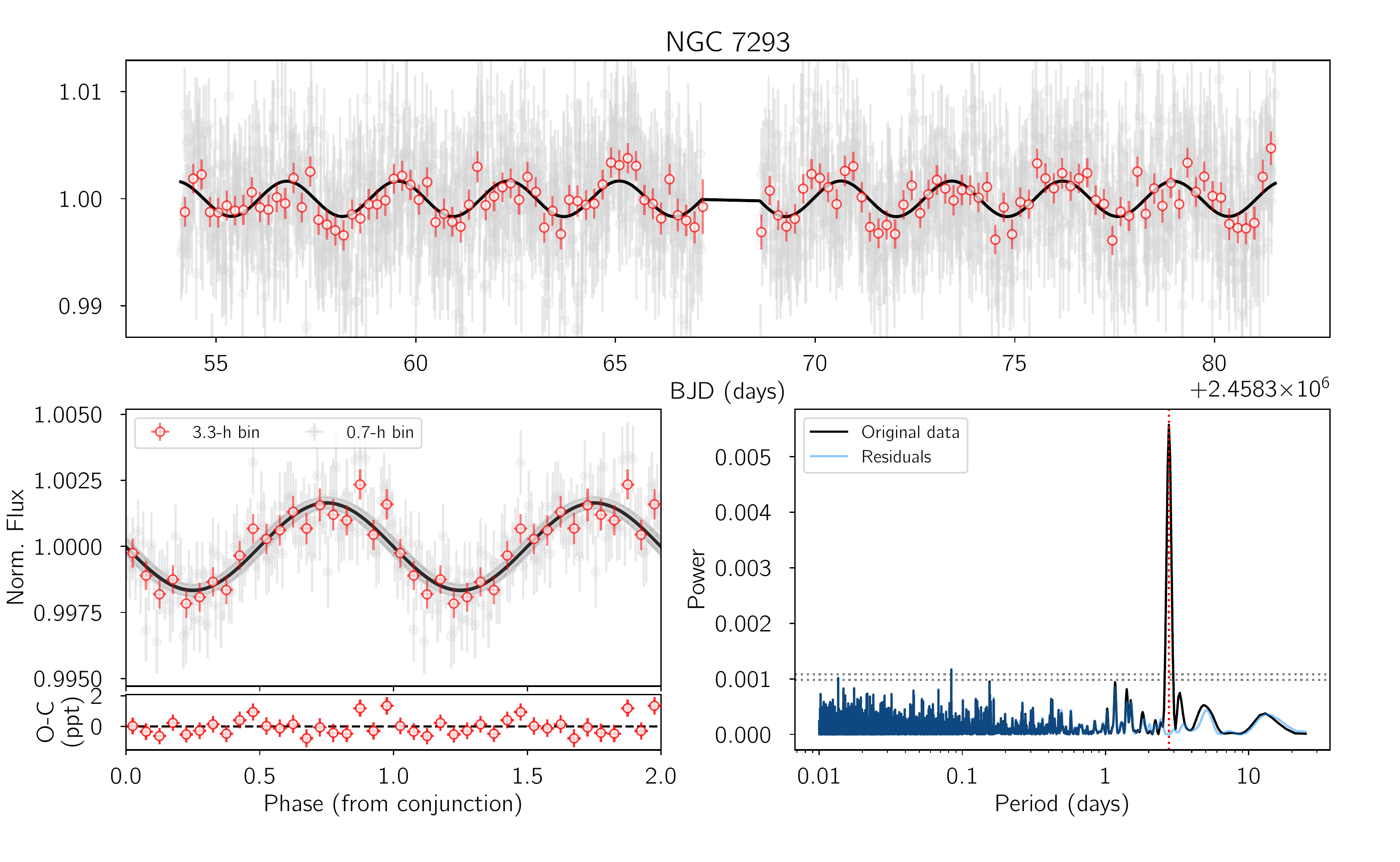}
        \includegraphics[width=0.99\textwidth]{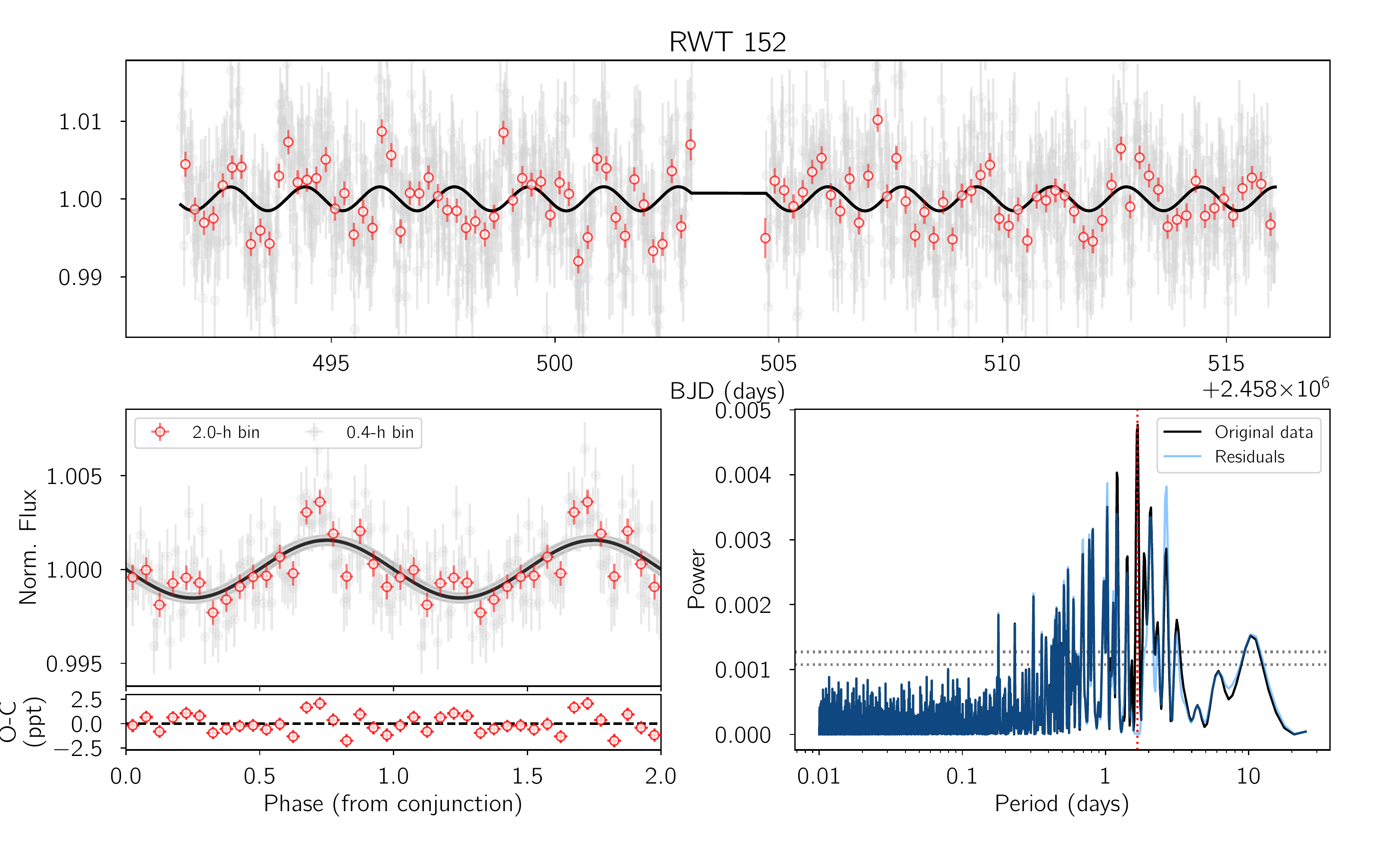} 
      \caption{Time-flux light curves (top panels), phase-folded light curves (left panels) and periodograms (right panels) of NGC\,7293 and RWT\,152. Two different bin sizes are shown with red and grey circles (see legend for each individual object). The periods derived from the corresponding fittings can be found in Table~3. }
\label{fig:lightcurves_and_periodogram1}
\end{figure*}

 The periodogram of the TESS light curve is quite different from the rest of the central stars in the sample (see Fig.~\ref{fig:lightcurves_and_periodogram1}, second panel). We find a forest of significant periodicities above the 0.1\% FAP between 0.1 and 10 days approximately. The MCMC analysis found the most prominent peak (corresponding to $\sim$ 1.7 d period) as the most probable, with an amplitude of 1.6 ppt (similar to that found in NGC\,7293). The nature of the most significant periodicity as well as the other significant peaks in the periodogram is not easily interpretable and radial velocity observations are essential for unveiling it.

 One possibility is that the observed variability is caused by pulsations or stellar rotation. Thanks to a frequency analysis in the time domain \citep{BarceloForteza2015}, we found the most significant peak at 6.2 $\mu$Hz (or 1.9 days) and its closest peak of similar amplitude at 5.6 $\mu$Hz (or 2.1 days). Both peaks may correspond to the rotation period of a star with differential rotation. We also see multiples of this signature such as twice ($\sim$11 $\mu$Hz) and thrice ($\sim$17 $\mu$Hz) its value (see Table~\ref{t:multi}) with a non-significant deviation in frequency, lower than the resolution (7\%). These signatures may be produced by the presence of spots. In addition, we used the wavelet transform \citep{Mathur2010} to find a proper value of rotation, obtaining similar results: 2.0 $\pm$ 0.6 days (5.8 $\pm$ 1.6 $\mu$Hz).\\

\begin{table}
\caption{Multiples of rotational peaks of the power spectrum of RWT\,152.}
\label{t:multi}
\centering
\begin{tabular}{c | c c}
\hline\hline
Multiple & Frequency ($\mu$Hz)  & Deviation (\%)\\
\hline
1  & 6.189 $\pm$ 0.017  & -  \\
1  & 5.616 $\pm$ 0.017 & -  \\
1  & 5.139 $\pm$ 0.022 & -  \\
2  & 11.257 $\pm$ 0.018  & 0.4  \\
3  & 16.992 $\pm$ 0.023  & 2.5  \\
4  & 22.424 $\pm$ 0.025  & 0.8  \\
5  & 26.035 $\pm$ 0.025  & 6.6    \\
6  & 36.848 $\pm$ 0.025  & 4.6    \\
 \hline 
\end{tabular}
\end{table}


   \begin{figure*}
        \includegraphics[width=0.99\textwidth]{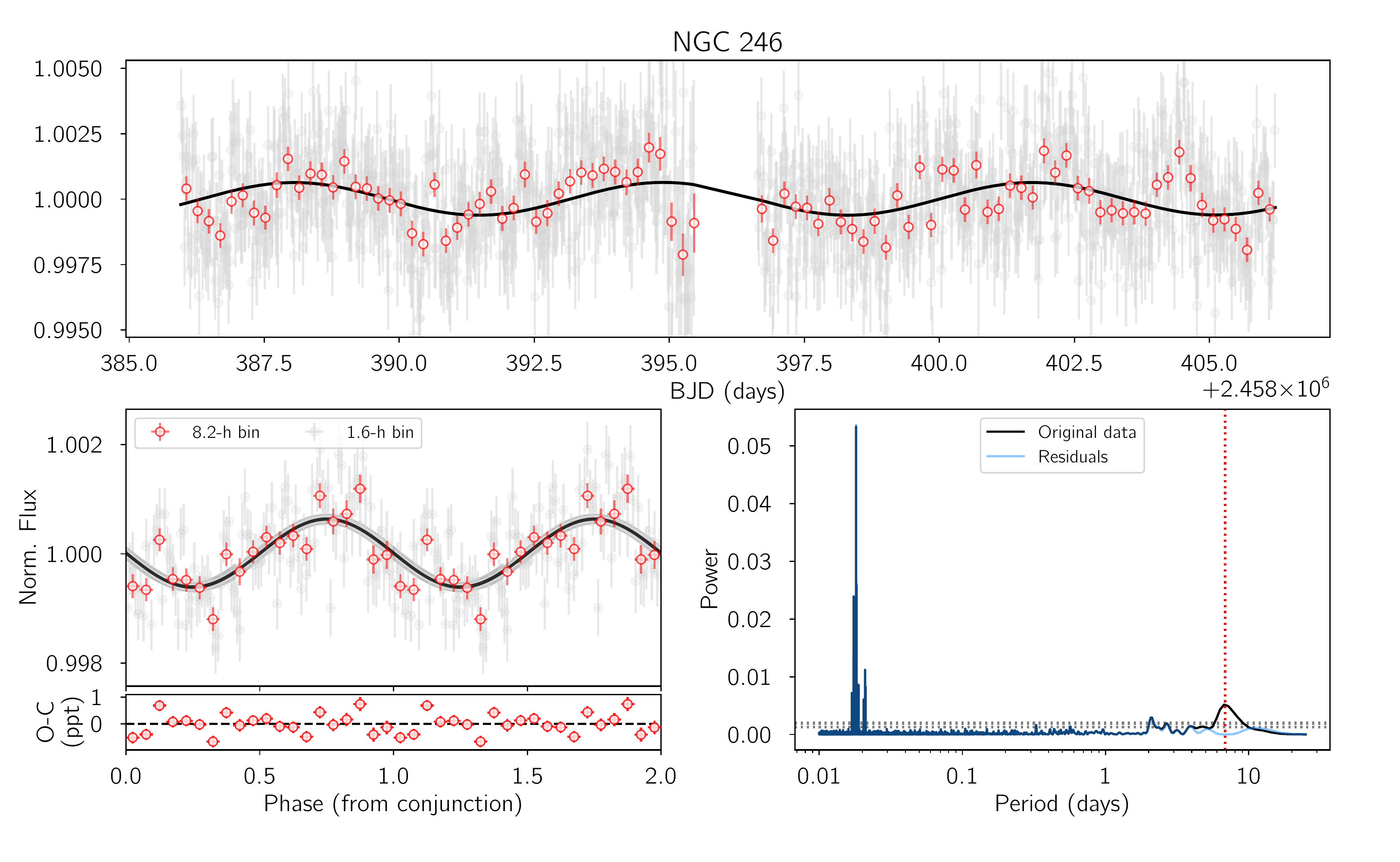}
         \includegraphics[width=0.99\textwidth]{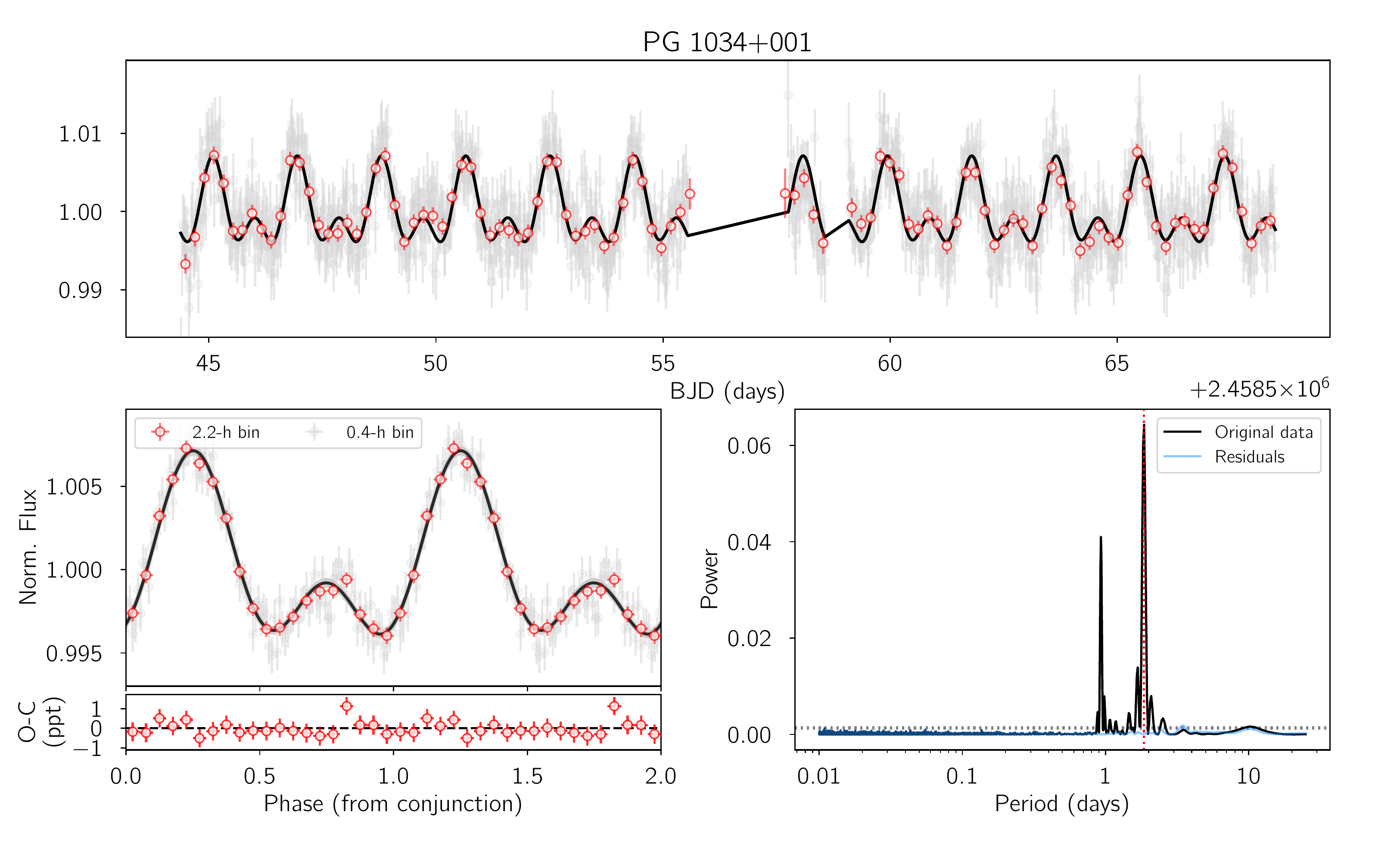}                         
  \caption{Time-flux light curves (top panels), phase-folded light curves (left panels) and periodograms (right panels) of NGC\,246 and PG\,1034+001. Two different bin sizes are shown with red and grey circles (see legend for each individual object). The periods derived from the corresponding fittings can be found in Table~3 (for NGC\,246) and Table~4 (for PG\,1034+001).}
\label{fig:lightcurves_and_periodogram2}
\end{figure*}

   \begin{figure*}                        
        \includegraphics[width=0.99\textwidth]{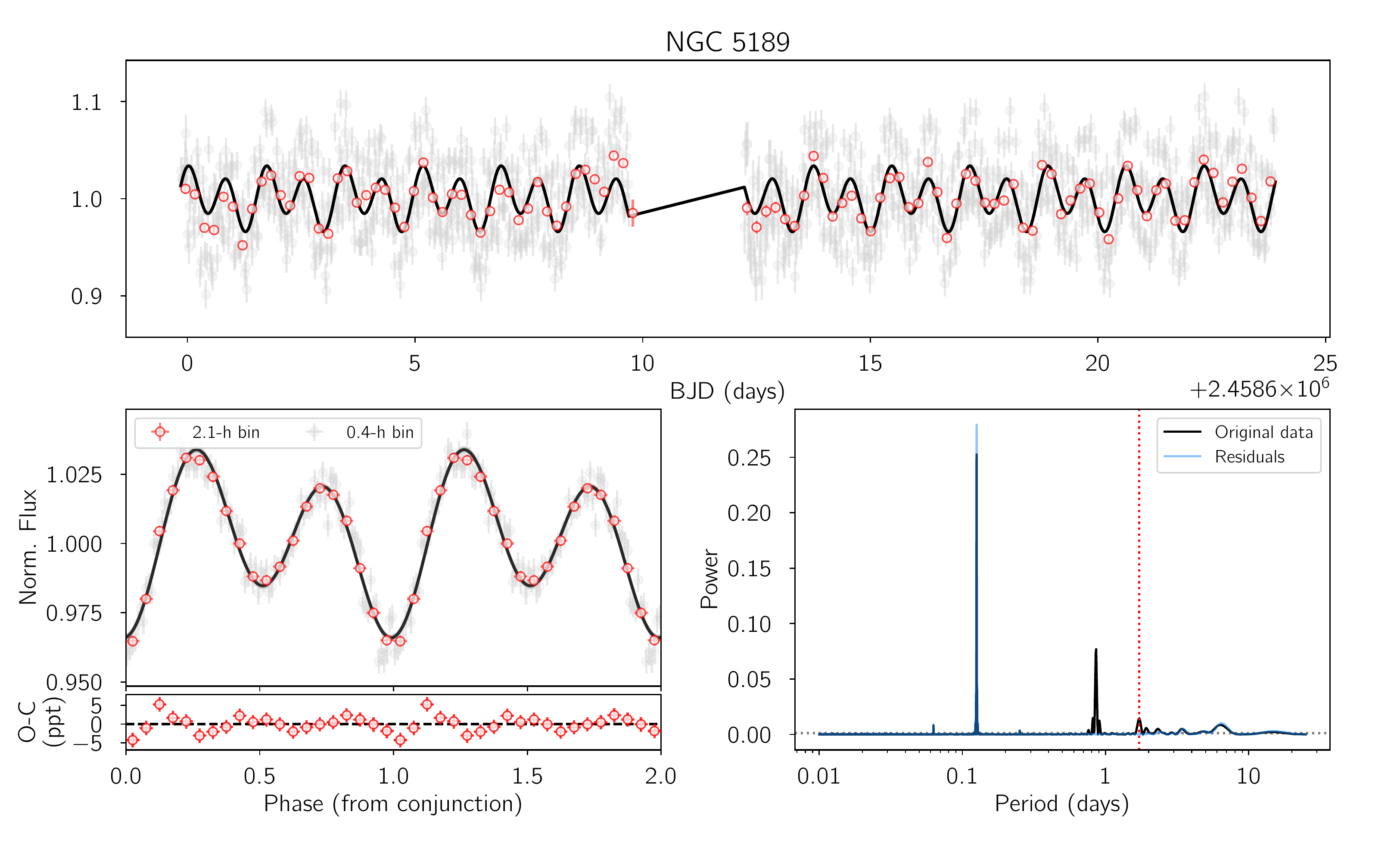}   
                  \includegraphics[width=0.99\textwidth]{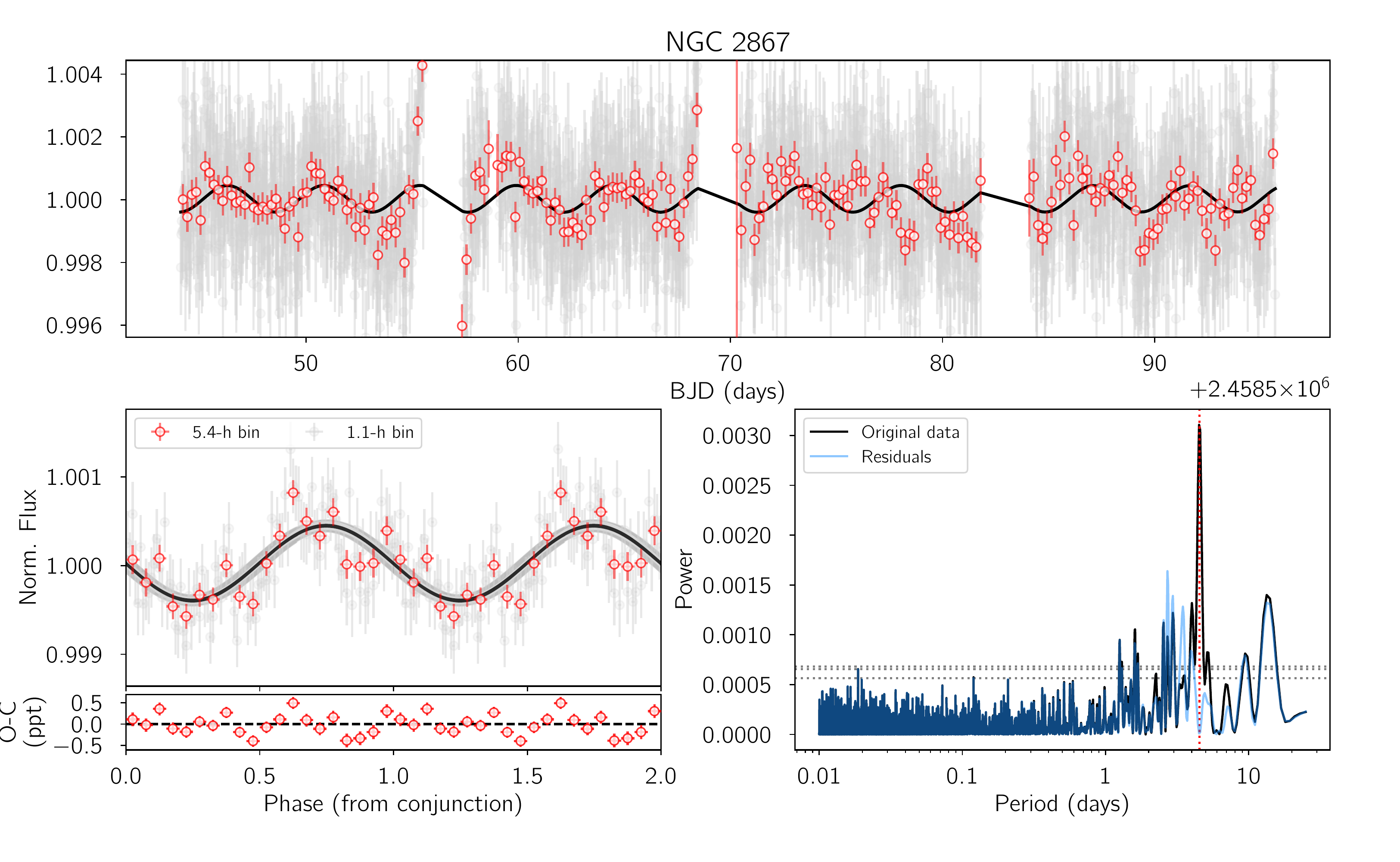}
    \caption{Time-flux light curves (top panels), phase-folded light curves (left panels) and periodograms (right panels) of NGC\,5189 and NGC\,2867. Two different bin sizes are shown with red and grey circles (see legend for each individual object). The periods derived from the corresponding fittings can be found in Table~3 (for NGC\,2867) and Table~4 (for NGC\,5189).}
    \label{fig:lightcurves_and_periodogram3}
\end{figure*}

   \begin{figure*}
       \includegraphics[width=0.99\textwidth]{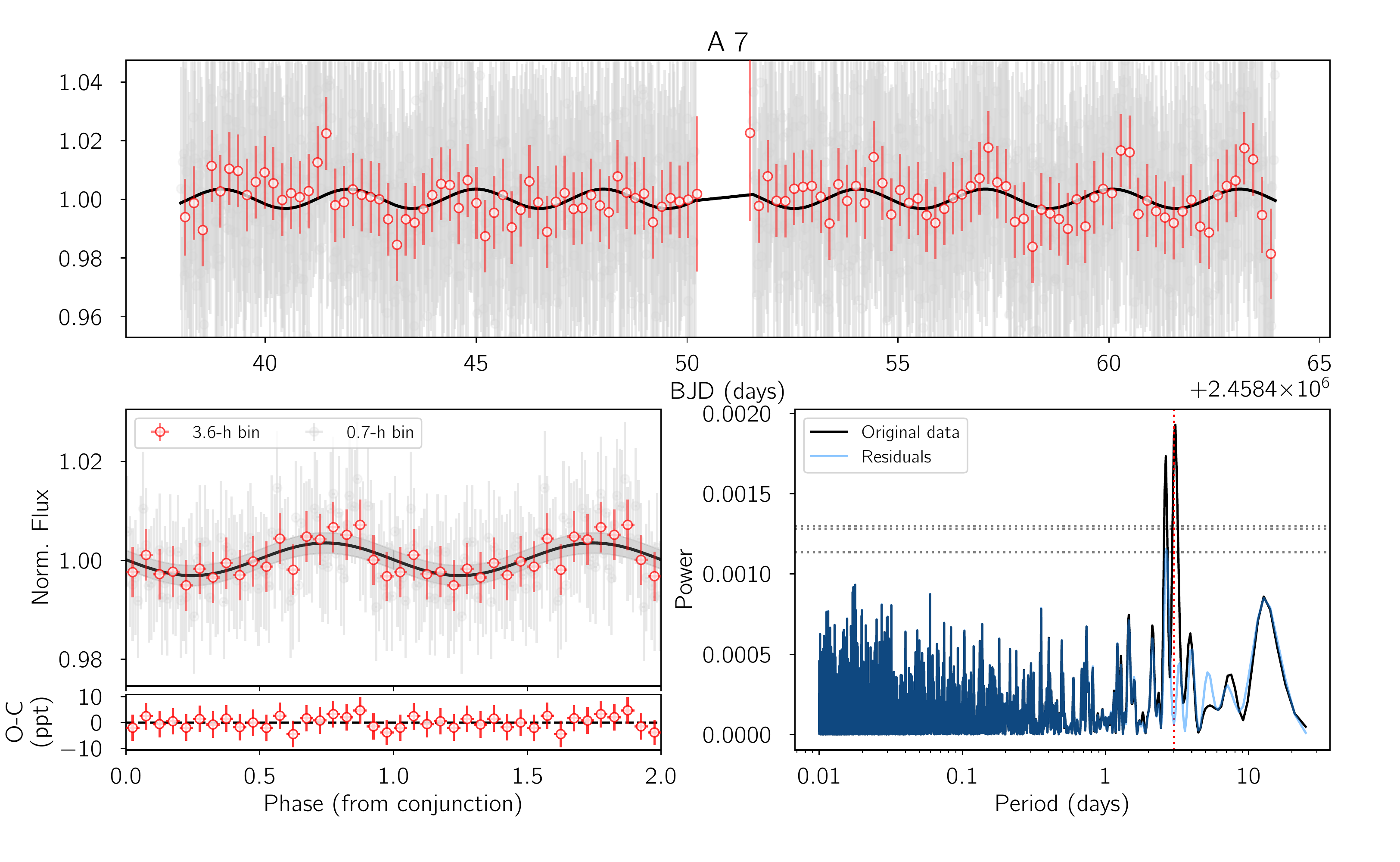}
        \includegraphics[width=0.99\textwidth]{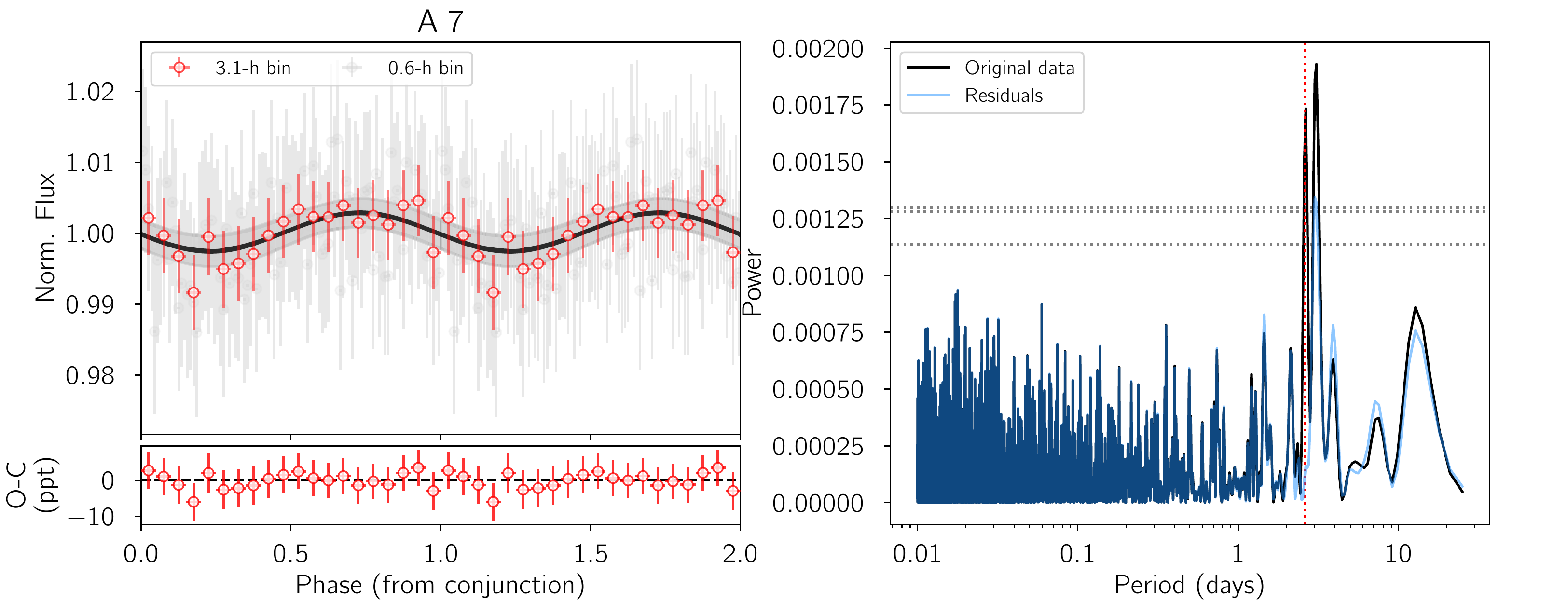}  
  \caption{Time-flux light curve (top panel), phase-folded light curves (left panels) and periodograms (right panels) of Abell\,7. Two different bin sizes are shown with red and grey circles.The periods derived from the corresponding fittings can be found in Table~3.}\label{fig:lightcurves_and_periodogram4}
\end{figure*}

   \begin{figure*}
               \includegraphics[width=0.99\textwidth]{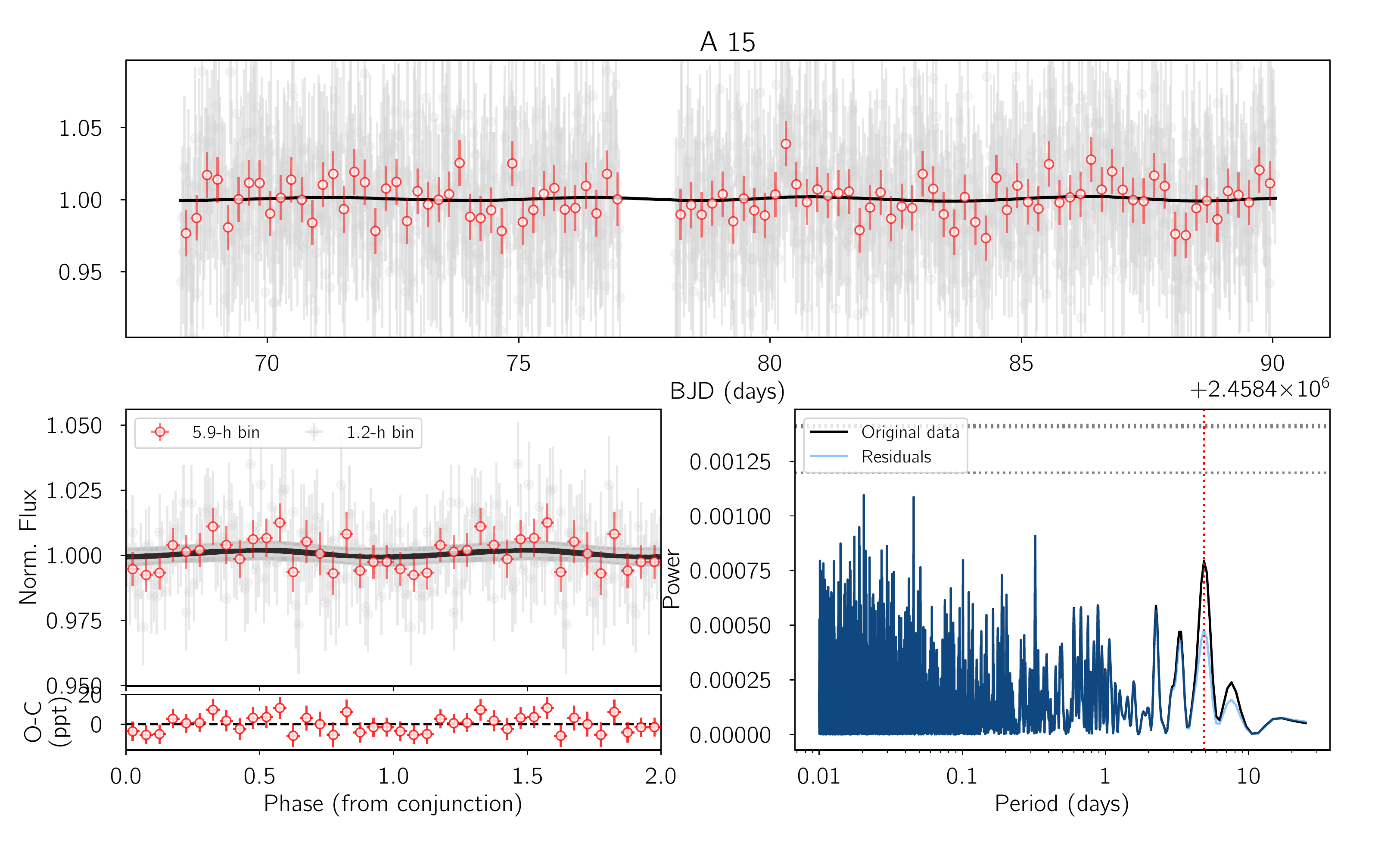}
  \caption{Time-flux light curve (top panel), phase-folded light curve (left panel) and periodogram (right panel) of Abell\,15. Two different bin sizes are shown with red and grey circles.}
\label{fig:lightcurves_and_periodogram5}
\end{figure*}


   \begin{figure*}
       \includegraphics[width=0.99\textwidth]{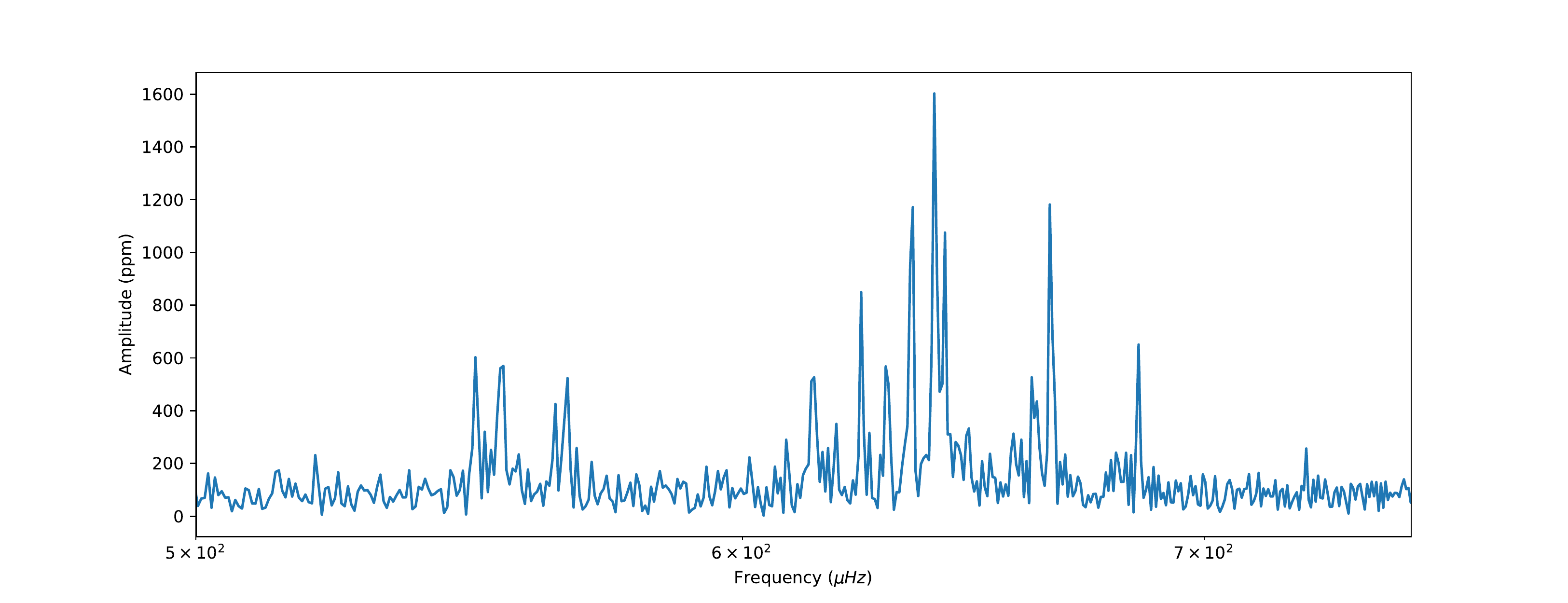}
      \caption{Periodogram of NGC\,246 in the low frequencies. The different peaks are due to the pulsations.}
\label{fig:pulsations_ngc246}
\end{figure*}

\subsection{NGC\,246}

NGC\,246 exhibits an elliptical, shell-like appearance with a clumpy structure. The western edge is considerably brighter than the rest of the shell. This, and the displacement of the PN nucleus from the geometrical center of the nebula support the idea of the interaction with the ISM \citep[for more details, see e.g.][]{Muthu2000,Szentgyorgyi2003}. The central star of this PN (HIP 3678) is a known hierarchical triple stellar system with an early to mid-K dwarf (K2-K5, named HIP 3678 B) and a mid M dwarf (M5-M6, named HIP 3678 C) as companions of the primary white dwarf \citep[see][]{Adam2014}. This hot, hydrogen-deficient planetary nebula nucleus (PNN) was also identified as a low-amplitude pulsator with periods in the range of 24-31 min (i.e. 543-683  $\mu$Hz) and amplitude of $\sim$2 mmag \citep{Ciardullo-Bond1996}. Later, \cite{GonzalezPerez2006} found significant daily changes in these pulsations, with the appearance of a new peak in the power spectra, which is peculiar in comparison with other observed PNNi, which have some variations in the amplitudes of certain peaks but never the appearance of a new peak. 

In the TESS light curve we clearly identify the periodic photometric variability due to the aforementioned pulsations (with several significant peaks in the periodogram from 550 to 700 $\mu$Hz approximately, see Fig.~\ref{fig:pulsations_ngc246}). For the first time, we are able to resolve all the individual frequencies due to the different pulsations modes. We can now conclude that what previous authors observed was, in fact, the broad envelope corresponding to all the 'forest'\ of frequencies in the periodogram, which are concentrated in two different distributions in the periodogram. However,  \cite{GonzalezPerez2006} and \cite{Ciardullo-Bond1996} did not have enough precision to resolve all the individual frequencies. We reserve a more detailed analysis of these pulsations for a future publication (Sowicka et al., in prep.).

In addition to the frequencies associated with pulsations, we found a new periodic photometric variability with a period of 6.8 days and an amplitude of nearly 0.6 ppt (see Fig.~\ref{fig:lightcurves_and_periodogram2}, first panel). This variability may be ascribed to a new source in the system. It is important to note that, as mentioned above, the photometry of this central star could be contaminated by another source that is inside the aperture mask. This source is HIP 3678 B and has a Gaia magnitude of G = 14.2 (2.4 magnitudes fainter than the central star HIP 3678 A).  Both stars are in the same pixel of the TESS camera, which makes impossible to separate the contribution of both sources. The difference in magnitudes of the two stars corresponds to a $\sim$11\% of contamination of the fainter star over the brighter. Hence, the origin of the variability is not clear and dedicated ground-based photometry on the two sources is necessary for unveiling its origin.

  \begin{table*}
 \centering
  \caption{Results from the light curve study for the five central stars analyzed with a simple sinusoidal model. }
    \begin{tabular}{lccc}

  \hline    
Object   &      Period (days)  &  $T_0$ (2400000) & Amplitude (ppt)\\
  \hline \hline
NGC\,246                &       6.8373$^{+0.1407}_{-0.1456}$ &  58355.5673$^{+0.8618}_{-0.8380}$        &         0.6327$^{+0.0704}_{-0.0705}$ \\
NGC\,7293       &       2.7748$^{+0.0169}_{-0.0166}$ &  58363.0140$^{+0.0564}_{-0.0573}$        &         1.6782$^{+0.1823}_{-0.1810}$ \\
NGC\,2867       &       4.5350$^{+0.0242}_{-0.0246}$ &  58361.5024$^{+1.0981}_{-1.0928}$        &         0.4283$^{+0.0480}_{-0.0459}$ \\
RWT\,152                &       1.6682$^{+0.0077}_{-0.0076}$ &  58379.7239$^{+0.5659}_{-0.5772}$        &         1.5843$^{+0.2065}_{-0.2083}$ \\
Abell\,7 (P1)   &       3.0172$^{+0.0101}_{-0.0100}$ &  58367.3184$^{+0.3749}_{-0.3784}$        & 3.5734$^{+1.7102}_{-1.7710}$ \\
Abell\,7 (P2)   &       2.6085$^{+0.0385}_{-0.0340}$ &  58369.4441$^{+2.4420}_{-2.3169}$        & 3.0886$^{+1.7751}_{-1.7420}$ \\
\hline
\end{tabular}
\label{Table:results_LCs}
\end{table*}

      \begin{table*}
 \centering
  \caption{Results from the light curve study for the two central stars analyzed with models including reflection, ellipsoidal and doppler beaming effects.}
    \begin{tabular}{lcc}

  \hline    
Parameter                       & NGC\,5189                                     &       PG\,1034+001\\
  \hline \hline
Period (days)                   &       1.7159$^{+0.0007}_{-0.0007}$            &  1.8566$^{+0.0015}_{-0.0015}$ \\
$T_0$ (2450000)                 &       58367.9266$^{+0.1060}_{-0.1025}$                 &  58362.6786$^{+0.1528}_{-0.1538}$ \\
A$_{ell}$ (ppt)         &       25.6353$^{+0.5945}_{-0.6215}$           &  3.1593$^{+0.1487}_{-0.1534}$ \\
A$_{ref}$ (ppt)         &       9.4441$^{+0.6125}_{-0.6132}$            &  0.0989$^{+0.1151}_{-0.0710}$ \\
A$_ {beam}$ (ppt)       &       6.6951$^{+0.6174}_{-0.6432}$            &  3.9626$^{+0.1517}_{-0.1516}$ \\
\hline
\end{tabular}
\label{Table:results_LCs2}
\end{table*}

\subsection{PG\,1034+001}
\label {pg1034+001}

The PN surrounding this hot hydrogen-deficient DO white dwarf is probably the largest known PN on the sky. It was discovered by using the Sloan Digital Sky Survey spectra and confirmed with narrow-band H$\alpha$ and [O\,{\sc iii}] images, which revealed a complex and non-defined nebulosity with an angular size of about 2$^\circ$ of diameter \citep{Hewett2003}. Later, \cite{Rauch2004} reported the detection of other much more extended emission structures around this central star.

In the Lomb-Scargle periodogram of the TESS dataset (see Fig.~\ref{fig:lightcurves_and_periodogram2}, second panel), we find two prominent peaks above the 0.1\% FAP, at 1.85 d and 0.93 d. We can definitively state that the signal at 0.93 d must be an alias, since when we recalculate the residual Lomb-Scargle periodogram after subtracting the signal at 1.85 d, the other peak at 0.93 d is no longer present. Fig.~\ref{fig:lightcurves_and_periodogram2} shows the light curve phase-folded with the strongest period of 1.85 d. At first glance, it is possible to interpret the variability as the result of at least two different effects. On the one hand, we can identify a sinusoidal curve with two maxima per orbital period, which points to ellipsoidal modulation variability. This is the imprint in the light curve of the tidal distortions caused by the gravitational interaction between the two components of a binary system: the secondary star acts as a gravitational perturber to the external layers of the primary, inducing changes in the spherical shape of the star resulting in variability in the light curve along the orbital path. On the other hand, the asymmetry in the ellipsoidal modulation could indicate the presence of relativistic Doppler-beaming effects \citep{Zucker2007}. This effect is basically the photometric signature of the radial velocity variations. The shifted spectral energy distribution due to the velocity variations produces flux variability as seen by a wavelength-fixed bandpass. Its detection from ground based telescopes is challenging because of the low amplitudes that this effect produces ($\le$0.1\%; Zucker et al. 2007) and the high precision and continuous observations along a long-timespan needed to detect it. To date, the only bCSPN exhibiting Doppler- beaming effects is the central star of the multi-shell PN AMU\,1 \citep{Aller2013}, found in the analysis of the Kepler light curves by \cite{DeMarco2015}. First Kepler and now TESS have opened the window to this new detection technique of stellar variability, in particular in the field of bCSPNe. Finally, irradiation effects (i.e. the reflection of the light of the hot primary by a less massive companion) could also be present. In the phase-folded light curve, the final model is a combination of the equations for the three effects (reflection, ellipsoidal variability, and Doppler beaming). The amplitudes of each effect are listed in Table~4. We can conclude that the predominant effect would be Doppler beaming and that reflection is the least significant. However, some incongruities can be derived from these values, especially from the amplitude of the Doppler beaming, which we will discuss in the following. 
  
It is interesting to compare the light curve of PG\,1034+001 to that of the binary star KOI-74, which consists of an A-type star and a white dwarf \citep{Bloemen2012,vanKerkwijk2010}. The main difference between both light curves is that apart from reflection, ellipsoidal modulation, and Doppler beaming, KOI-74 also shows eclipses in its light curve. In the case of PG\,1034+001, the amplitude of the possible Doppler beaming variations is a factor $\sim$ 30 larger than the same amplitude in KOI-74. Following equation (2) we can estimate the radial velocity amplitude of the star from the Doppler beaming (as we did in the case of NGC\,7293). Assuming a temperature of T$_{\rm eff}$ = 115000\,K for the primary \citep{Werner2017}, we derive $f_{\rm DB}$ = 1.08, which corresponds to a radial-velocity semi-amplitude of $\varv{_r}$ $\approx$ 1095 km s$^{-1}$. This yields a mass function of $f$(m) = 252 M$_{\sun}$. With one of the components being a white dwarf \citep[with 0.62 M$_{\sun}$ determined by][]{Werner2017} this would imply a minimum mass for the companion of $\approx$ 252 M$_{\sun}$, essentially ruling out Doppler beaming.

Another more plausible scenario to account for such a huge amplitude is that the variability we are detecting is the consequence of the O'Connell effect \citep{oconnell51,milone68, Johnston2019}, whereby ellipsoidally-modulated binaries are frequently found to present uneven maxima in their light curves.
It had previously been suggested that this could be the result of a so-called `periastron effect' in which the binaries are somewhat eccentric and experience an increased irradiation effect close to periastron \citep{roberts06}; however it quickly became clear that in most cases the brighter maxima does not coincide with periastron \citep{dugan16} or that their orbits were essentially circular \citep{oconnell36}, thus ruling out this hypothesis.  It is now generally thought that the effect is likely due to a hot or cool spot (or spots) on one of the component stars, perhaps via accretion streams which would naturally lead to hot spots \citep{wilsey09}.  However, introducing such spots into a model leads to significant degeneracy, making the drawing of reasonable conclusions nigh on impossible without employing Doppler tomography techniques \citep{bell90}.  As such, we make no attempt to characterise the nature of PG\,1034+001, which appears to display a pronounced O'Connell effect.

While a binary origin is the most plausible for the variability shown by PG 1034+001, it is important to note that spots on a single star could also reproduce the observed light curve, particularly given the recent observational evidence for the existence of long-lived spots on white dwarfs and similar objects (see e.g. Dupuis et al. 2000; Kilic et al. 2015).

\subsection{NGC\,5189}

NGC\,5189 exhibits a very complex morphology, with multiple filamentary and knotty structures distributed over the nebula \citep[see e.g.][]{Danehkar2018}. \cite{Sabin2012} classified it as a quadrupolar PN and also reported the detection of a dense and cold infrared torus. The nucleus of NGC\,5189 is a [WO1] Wolf-Rayet-type CS and was recently identified as a binary by \cite{Manick2015} with a period of 4.04 d. They also found other significant peaks at  3.54, 1.32, and 0.80 d, although they concluded that these are most likely to be aliases. In that paper, the authors proposed as the most plausible companion a massive white dwarf with a mass of m$_2$ $\ge$ 0.5 M$_\sun$.

In Fig.~\ref{fig:lightcurves_and_periodogram3}, we present the periodogram of the TESS data, which shows several peaks above the 0.1\% FAP. 
Before computing the MCMC fitting, we visually inspected the most prominent peaks individually, finding several sources of periodicities. On the one hand, the strongest peak at 0.125 d shows a clear variability with $\sim$ 40\,ppt amplitude (see Fig.\,\ref{appfig:ngc5189}), which we were unable to properly fit with a simple sinusoidal function. We also tried the three-component model mentioned previously, but the results provided an unphysical solution for the amplitude of the beaming effect. The alias of this signal is also significant in the periodogram at 0.062 d and it does not persist in the residual Lomb-Scargle periodogram once the first periodicity is subtracted (with the three-component model). Neither of the periodicities identified result in a phase-folded light curve which can be easily associated with typical binary variability.  However, in spite of the strong deviations from the sinusoidal, the light curve when folded on the strongest peak in the periodogram (see Fig. A.1) could feasibly be the result of irradiation in a high inclination system (which can result in strong asymmetries between minima and maxima).

The second most significant and interesting peak is 0.858 d, which shows a sinusoidal function when we phase-folded the light curve with such period. However, after inspecting the signal at 1.71 d (twice the period of 0.858 d), we realised that this was the true period since we have two minima per cycle. As the case of PG\,1034+001, we identify three possible effects in the light curve of NGC\,5189: reflection, ellipsoidal variability or Doppler beaming. In Fig.~\ref{fig:lightcurves_and_periodogram3}, we have plotted the light curve phase folded with this period and the final model accounting for the three effects. In the residual periodogram can be seen that the strong peak at 0.858 d is an alias. However, the 0.125 d period (and its alias)  persist in the periodogram, which means that the nature of these two periodicities are different.  As in the case of PG\,1034+001, the amplitude of the Doppler beaming derived from the fitting of the light curve is too large and would imply a very massive companion in the system (see Sect. ~\ref{pg1034+001} for more details), which is somewhat difficult to explain. Again, the most plausible and likely explanation is that the asymmetry in the ellipsoidal modulation is due to a star spot and so, we are indeed detecting the O'Connell effect and not the Doppler beaming. We speculate that a plausible scenario would be a system with two white dwarfs, as proposed by \cite{Manick2015}, but with different effective temperatures. However, the lack of information about the system does not allow us to do a more detailed model of the binary and radial velocity observations are required to constrain the parameters of the two components.

Surprisingly, we do not see any variability at the 4.04-day period previously published by \cite{Manick2015} to be the strongest in their radial velocity data. Among the other periodicities mentioned by the authors, we find photometric counterparts close to their 3.54 d and 0.80 d periods (the corresponding peaks in the TESS periodogram are located at 3. and 0.85, respectively, both are above the FAP 0.1\,\%). Finally, we find no evidence for the 1.32 d periodicity, which has also been reported by  \cite{Manick2015}.

 It is really important to consider that, as mentioned in Sect.~2.2, the aperture mask used to extract the photometry of NGC\,5189 is subject to severe contamination by several sources of similar magnitudes as the central star (see Fig.\ref{fig:TPFs}). This makes it very difficult to discuss where both periodicities might come from.
Another possibility for overcoming the apparent discrepancy between the radial velocity data from \cite{Manick2015} and the photometric data presented here is the presence of stellar winds. It is known that Wolf-Rayet-type CSPNe have strong stellar winds with mass-loss rate up to $\sim$ 10$^{-7}$ M/yr.  Stellar wind variability may
introduce some apparent velocity variations in the absorption line profiles \citep[see e.g.][]{DeMarco2004}. Therefore, a contribution of a variable stellar wind could be present in the radial velocity curve of NGC\,5189 leading to the derivation of a (perhaps) spurious period where there is no sufficient data to separate the variability due to wind and orbital motion.

\subsection{NGC\,2867}

This PN is catalogued as round or elliptical with multiple shells or haloes by \cite{Stanghellini1993}. As the case of NGC\,5189, NGC\,2867 is a planetary nebula with a [WO1-2] spectral type as a central star. It was identified as a pulsating star by  \cite{GonzalezPerez2006}, with a period of 769s ($\sim$ 12 min). For this object, we found two different light curves in the TESS archive, taken in sectors 9 and 10, respectively. We inspected the target pixel file (TPF) of both and checked the Tess Input Catalogue (TIC) name and the coordinates to be sure that both light curves correspond to the same object. For the analysis, we combined both data sets into a single light curve.

We detect a significant periodic variability at 4.5 days and amplitude of 0.4 ppt (see Fig.~\ref{fig:lightcurves_and_periodogram3}). However, as in the case of NGC\,5189, the aperture mask includes several other stars, some of them with similar magnitude to our target, which makes it challenging to identify the origin of the variability. In this case, photometric ground-based follow-up, with better spatial resolution, of each target within the aperture is required in order to discern the source of the variability. We note that detections of photometric variations of 0.4\,ppt in amplitude are feasible from ground-based telescopes.

\subsection{Abell\,7}
\label{Sect: Abell7}

Abell\,7 is a very faint, slightly elongated PN (with a size of 15$'$$\times$13$'$) and a clumpy structure inside the main shell (see deep image by D. Goldman\footnote{\url{https://astrodonimaging.com/gallery/abell-7/}}). The central star of this PN was classified as a possible physical binary by \cite{Ciardullo1999} after resolving a very
faint companion ($\sim$ 5.3 mag fainter than the central star) 0.91 arcsec away in the HST frames. \cite{DeMarco2013} estimated an M5V spectral type for this wide companion based on the measurement of I- and J-band flux excess. 

In the analysis of the TESS light curve, we identify two clear strong peaks at approximately 2.6 days and three days. The periodogram of the residuals (in blue in Fig.~\ref{fig:lightcurves_and_periodogram4}) indicates that both signals are significant since after removing one of the signals, the other still remains above the 0.1\% FAP. Hence, both signals are independent and most probably the origin of each is different. In any case, the periods of the detected signals could not be compatible with an M star at 0.91 arcsec and, as consequence, both photometric periodicities would be associated with other companions (if they are indeed of binary origin).

\subsection{Abell\,15}

This old PN was discovered by \cite{Abell1966} on the Palomar Observatory Sky Survey (POSS) plates. The nebula has been classified as round or ring-shaped PN \citep{Phillips2003, Hromov1968}. However, the lack of deep images or high-resolution, long-slit spectra of the nebula prevent us from doing a more detailed description of the morphology. Its central star is the only one in our sample that does not have significant peaks in the periodogram (see Fig.~\ref{fig:lightcurves_and_periodogram5}). The light curve does not show variability above $\sim$ 8 mmag at the 95\% confidence level.  This non-detection of variability is not surprising since Abell\,15 is, apparently, the only round PN in our sample (only Abell\,7 would be close to round, although it shows a slightly deviation from sphericity; see above). This result is compatible with the idea that
round PN are formed from single stars.

\section{Final remarks and conclusions}

      We analysed the TESS light curves available in the MAST archive of eight CSPNe in order to identify new binary candidates based on high-precision photometry. In all cases except one, we detect clear signs of variability, which means a positive detection rate of more than 80\%. In most of the cases, these photometric variations can be attributed to the effect of a companion star or stars to the nebular progenitor. In seven out of the eight objects, the light curves show a variability that is consistent with an irradiation effect. Two of the targets (PG\,1034+001 and NGC\,5189) also exhibit ellipsoidal variability. We also discuss the possible detection of relativistic Doppler beaming effects and the O'Connell effect in several objects. In the case of NGC\,7293 (Helix Nebula), the most well-studied PN in the sample, it was possible to construct a series of binary models to probe the possible companion properties which would lead to the observed variability in its light curve. These models lead us to conclude that if the variability is, indeed, due to binarity then the companion is likely a very low-mass main-sequence star or sub-stellar object.
        
               The orbital periods found for the seven central stars range from 1.6 to 6.8 days, approximately. These periods would imply that the systems are post-common-envelope and while they are not exceptionally long, they do lie in the longer period tail of the observed post-common-envelope bCSPN period distribution. 
               
               In the case of the central star of NGC\,246, an already known low- amplitude pulsator, the signals attributed to the pulsations are clearly identified in the periodogram. For the first time, all the frequency modes are resolved though we reserve a full analysis of these modes for a future paper (Sowicka et al., in prep.). 
               
      The only central star for which no significant variability is found is Abell\,15. This is the only PN in the sample with an apparently round morphology, which is consistent with the theory that round PNe are formed via a single star evolution. Intriguingly, the other seven detections are PNe with more complex morphologies, which are believed to be typical of central star binarity.
      
      Three central stars in the sample exhibit a possible contamination by other sources in the TESS aperture within a magnitude limit of $\Delta$m = 3. For these, ground-based photometry is necessary in order to discern the source of the variability. Also, radial velocity follow-up is essential in all cases to constrain the orbital parameters of the binary candidates found in this work.
     
\begin{acknowledgements}

We thank our anonymous referee for their useful comments that have improved the interpretation and discussion of the data. AA acknowledges support from Government of Comunidad Aut\'onoma de Madrid (Spain) through postdoctoral grant `Atracci\'on de Talento Investigador' 2018-T2/TIC-11697. JLB and SBF acknowledge support from the Spanish State Research Agency (AEI) through projects No.ESP2017-87676-C5-1-R and No. MDM-2017-0737 Unidad de Excelencia ``Mar\'ia de Maeztu''- Centro de Astrobiolog\'ia (INTA-CSIC). DJ acknowledges support from the State Research Agency (AEI) of the Spanish Ministry of Science, Innovation and Universities (MCIU) and the European Regional Development Fund (FEDER) under grant AYA2017-83383-P.  DJ also acknowledges support under grant P/308614 financed by funds transferred from the Spanish Ministry of Science, Innovation and Universities, charged to the General State Budgets and with funds transferred from the General Budgets of the Autonomous Community of the Canary Islands by the Ministry of Economy, Industry, Trade and Knowledge. LFM acknowledges partial support by grant AYA2017-84390-C2-R, co-funded with FEDER funds, and financial support from the State Agency for Research of the Spanish MCIU through the "Center of Excellence Severo Ochoa" award for the Instituto de Astrof\'isica de Andaluc\'ia(SEV-2017-0709). Authors are very grateful to Cristina Rodr\'iguez L\'opez for guiding us with the analysis of RWT\,152 and NGC\,246. This research has made use of the SIMBAD database, operated at the CDS, Strasbourg (France), Aladin, NASA's Astrophysics Data System Bibliographic Services, and the Spanish Virtual Observatory (http://svo.cab.inta-csic.es) supported from the Spanish MICINN/FEDER through grant AyA2017-84089. This publication makes use of data collected by TESS mission.
Funding for the TESS mission is provided by the NASA Explorer Program.
We acknowledge the use of pipelines at the TESS Science Office and at the TESS Science
Processing Operations Center. This paper makes use of data products from the Mikulski Archive for Space Telescope.

\end{acknowledgements}


\bibliographystyle{aa} 
\bibliography{/Users/Alba/Biblio_papers/Bibliography.bib} 

\Online

\begin{appendix} 
\section{Figures}

\begin{figure*}
\centering
\includegraphics[width=16.4cm,clip]{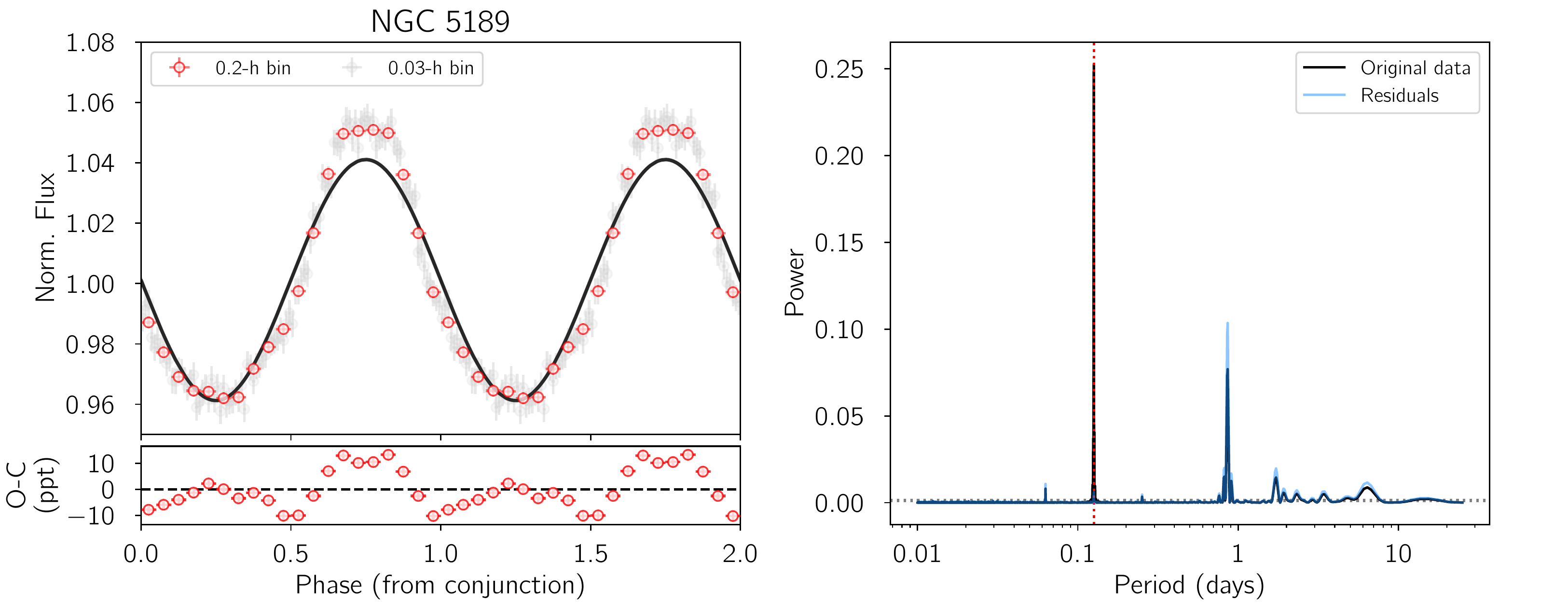}
\caption{Phase-folded light curve (left panel) and periodogram (right panel) of NGC\,5189.  Two different bin sizes are shown with red and grey circles.}
\label{appfig:ngc5189}
\end{figure*}

\end{appendix}

\end{document}